\documentclass[aps, prd, superscriptaddress, preprintnumbers]{revtex4}

\usepackage{graphicx,amsfonts,amsmath,amssymb,amstext}
\usepackage{float,wrapfig}
\usepackage{subfigure, psfrag}
\usepackage{dsfont}
\usepackage{color}
\usepackage{bm}

\newcommand{\be}{\begin{equation}}
\newcommand{\ee}{\end{equation}}
\newcommand{\ba}{\begin{eqnarray}}
\newcommand{\ea}{\end{eqnarray}}

\definecolor{purple}{rgb}{0.8,0,0.6}
\definecolor{darkgreen}{rgb}{0.00,0.6,0.00}

\begin{document}

\title{Inter-node superconductivity in strained Weyl semimetals}
\date{December 20, 2018}

\author{P.~O.~Sukhachov}
\affiliation{Department of Applied Mathematics, Western University, London, Ontario, Canada N6A 5B7}

\author{E.~V.~Gorbar}
\affiliation{Department of Physics, Taras Shevchenko National Kiev University, Kiev, 03680, Ukraine}
\affiliation{Bogolyubov Institute for Theoretical Physics, Kiev, 03680, Ukraine}

\author{I.~A.~Shovkovy}
\affiliation{College of Integrative Sciences and Arts, Arizona State University, Mesa, Arizona 85212, USA}
\affiliation{Department of Physics, Arizona State University, Tempe, Arizona 85287, USA}

\author{V.~A.~Miransky}
\affiliation{Department of Applied Mathematics, Western University, London, Ontario, Canada N6A 5B7}

\begin{abstract}
The effects of a strain-induced pseudomagnetic field on inter-node spin-triplet superconducting states in Weyl semimetals are studied by using the quasiclassical Eilenberger formalism. It is found that the Cooper pairing with spins parallel to the pseudomagnetic field has the lowest energy among the spin-triplet states and its gap does not depend on the strength of the field. In such a state, both electric and chiral superconducting currents are absent. This is in contrast to the superconducting states with the spins of Cooper pairs normal to the field, which support a nonzero chiral current and are inhibited by the strain-induced pseudomagnetic field. The corresponding critical value of the field, which separates the normal and superconducting phases, is estimated.
\end{abstract}

\maketitle

\section{Introduction}
\label{sec:Introduction}

The interplay of superconductivity (superfluidity) and topology has a long history
that goes back to the studies of superfluid $^3$He~\cite{Volovik:1988,Volovik-book:2003}.
One of the most interesting and sought after features of topological superconductors is the existence of the Majorana surface modes, which originate from the nontrivial topology of bulk states. Such modes may have important
applications in quantum computations, where the topological protection is invaluable for preventing decoherence and errors \cite{Nayak:2008}.
The increased interest to topological materials also drives a vigorous search for topological superconductors~\cite{Alicea:2012,Beenakker:2013,Stanescu:2013}.
(For reviews on topological superconductivity, see Refs.~\cite{Bernevig,Schnyder-rev:2015,Sato:2016evq}.)

Recently, a new class of three-dimensional (3D) topological materials, Weyl semimetals, was
discovered. These materials have a very unusual band structure with the valence and conduction bands touching
only at isolated points, known as the Weyl nodes, in the Brillouin zone. Even more importantly, their low-energy
quasiparticles are chiral fermions with a linear dispersion relation that are described by a relativisticlike
Weyl equation. The Weyl nodes can be viewed as the monopoles of the Berry curvature \cite{Berry:1984}
that carry nonzero topological charges. According to the Nielsen--Ninomiya theorem~\cite{Nielsen-Ninomiya-1,Nielsen-Ninomiya-2},
Weyl nodes in solids always occur in pairs of opposite chirality. In general, the pairs
of nodes can be separated by $2\mathbf{b}$ in momentum space and/or by $2b_0$ in energy. Note that while
$\mathbf{b}$, which is known as the chiral shift \cite{Gorbar:2009bm}, breaks the time-reversal (TR) symmetry,
$b_0$ breaks the parity inversion (PI). Because of their unusual topological properties,
Weyl semimetals quickly advanced to the forefront of condensed matter physics research (for reviews,
see Refs.~\cite{Yan-Felser:2017-Rev,Hasan-Huang:2017-Rev,Armitage-Vishwanath:2017-Rev}).

From the very beginning, Weyl materials were investigated as a possible platform for a
topologically nontrivial superconductivity. Generically, two distinctive types of superconducting pairing
of Weyl fermions could be considered \cite{Meng-Balents:2012,Cho-Moore:2012,Wei:2014vsa,Hosur:2014fba,Bednik:2015tha,Kobayashi2015,Kim-Gilbert:2016,Hashimoto2016}.
The first one is the \emph{inter-node} pairing of quasiparticles from the Weyl nodes of \emph{opposite}
chirality. The resulting state is sometimes referred to as a Bardeen--Cooper--Schrieffer (BCS)
ground state~\cite{BCS:1957}, which in the original context describes the pairing of the electrons with opposite spins.
By taking into account that there are several possibilities for
the inter-node pairing, including those in a spin-triplet channel, we will refrain from using the term
``BCS" in connection to such a pairing. The other possibility is the \emph{intra-node} pairing
that involves quasiparticles from the \emph{same} Weyl node. This leads to spin-singlet
Cooper pairs with nonzero momenta and results in a Larkin--Ovchinnikov--Fulde--Ferrell (LOFF) type ground state~\cite{LOFF:1964}.

The question regarding the type of pairing in the ground state is important and subtle.
It is possible that the inter-node pairing is energetically more favorable than the intra-node one \cite{Bednik:2015tha},
although the former has gapless nodes in the energy spectrum \cite{Meng-Balents:2012,Cho-Moore:2012} for Weyl superconductors.
It should be noted, however, that the outcome
of the energy competition might strongly depend on model details. For example, in the case of a simplified model
with a local interaction, the inter-node pairing with the vanishing momentum of Cooper pairs is disfavored \cite{Wei:2014vsa}.

An interplay of magnetic fields and superconductivity in Weyl materials is another interesting research topic.
While it is well known that magnetic fields inhibit superconductivity due to the Meissner effect, it was argued
that the situation may change in the limit of very strong fields \cite{Rasolt:1992zz,Rosenstein-Shapiro:2017}, where the quenching of
kinetic energy due to the formation of the Landau levels greatly assists the electron pairing. Still, the superconducting currents due to the
Meissner effect increase the energy of superconducting states and provide the backreaction on the magnetic field making the analysis extremely complicated.
On the other hand, in Weyl semimetals, there is an additional intriguing possibility to realize a pseudomagnetic
(or axial magnetic) field $\mathbf{B}_5$. Indeed, as was shown in Refs.~\cite{Zhou:2012ix,Zubkov:2015,Cortijo:2015jja,Cortijo:2016yph,Cortijo:2016,Pikulin:2016,Grushin-Vishwanath:2016,Liu-Pikulin:2016,Arjona-Vozmediano:2017,Arjona:2018ryu}, small mechanical strains in Weyl semimetals can be described by an effective axial
vector potential $\mathbf{A}_5$, which, unlike a usual electromagnetic gauge potential $\mathbf{A}$, is a directly
observable quantity, whose constant part can be interpreted as the chiral shift $\mathbf{b}$.
In some special cases of static deformations,
strains give rise to a pseudomagnetic field $\mathbf{B}_5=\bm{\nabla}\times \mathbf{A}_5$. (In contrast, as
we showed in Ref.~\cite{Gorbar:2018nuw}, this is not generically the case for multi-Weyl semimetals.)
Typical magnitudes of strain-induced pseudomagnetic fields range from about $0.3~\mbox{T}$ in the case of
static torsion \cite{Pikulin:2016} to about $15~\mbox{T}$ for bent samples \cite{Liu-Pikulin:2016}.
From the physical viewpoint, it is important that the pseudomagnetic field $\mathbf{B}_5$ couples to fermions from the Weyl nodes of
opposite chirality with different sign. This immediately leads to a distinct dynamics of Cooper pairs comparing to the case of a usual magnetic field.
Last but not least, due to the fact that the backreaction on the pseudomagnetic field is negligibly weak, the Meissner effect is absent for $\mathbf{B}_5$. It should be noted that some configurations of strain, such as a compressive strain \cite{Ruan:2016}, could overtilt Weyl nodes and allow for type-II Weyl semimetals \cite{Soluyanov:2015}. While the superconductivity in such materials could be enhanced at the critical value of the tilt (see, e.g., Refs.~\cite{Alidoust:2017,Li-Shapiro:2017} as well as a recent review on topological transitions in Ref.~\cite{Volovik:2018}), the underlying physics is different because the corresponding strains do not generate pseudomagnetic fields.

Recently, by using the quasiclassical Eilenberger approach \cite{Eilenberger:1968,Houghton-Vekhter:1998}, the authors of Ref.~\cite{Matsushita:2018bty} showed that the pseudomagnetic field could affect the superconducting states
and even induce nonzero charge and/or spin currents. By using a simplified model of a two-band superconductor, in
particular, it was found that the electric current is proportional to $\mathbf{B}_5$. In order to study the gap generation,
the authors employed the Ginzburg--Landau equation, where the gap appeared to be spatially modulated in the absence
of currents. The case of a more realistic Dirac semimetal was described briefly and the corresponding result suggested
that a nonzero spin supercurrent should be induced.

In this study, we investigate in detail the generation of gaps in the case of inter-node pairing in the minimal model of a TR symmetry broken Weyl
semimetal with two nodes separated by a nonzero chiral
shift. One of our principal findings is that the inter-node spin-triplet pairing with the spins of Cooper pairs parallel to the pseudomagnetic field is energetically most favorable.
On the other hand, the states with the spins of Cooper pairs normal to $\mathbf{B}_5$ are inhibited by the field.
In addition, we find that both electric and chiral supercurrents vanish for the energetically most favorable superconducting state.

The paper is organized as follows. In Sec.~\ref{sec:Model}, we introduce a simple model of a Weyl semimetal
with a broken TR symmetry and discuss how an external pseudomagnetic field couples to quasiparticles.
We also define the Bogolyubov--de Gennes (BdG) Hamiltonian and give its explicit form in the case of inter-node pairing. In Sec.~\ref{sec:Eilenberger-Derivation}, we present
the quasiclassical Eilenberger equation up to the second order in the spatial derivatives and pseudomagnetic
fields.
By using the iterative solutions of the Eilenberger equation, the superconducting gaps and currents are obtained and discussed in Sec.~\ref{sec:Eilenberger-Eqs-sol}.
The summary of the main results is given in Sec.~\ref{sec:Summary}. Technical details related to the derivation of the Eilenberger equation and its iterative solution are
presented in Appendices~\ref{sec:App-expansion} and \ref{sec:Eilenberger-sol-inter}, respectively. Throughout the paper, we use the units with $\hbar=c=k_{\rm B}=1$.

\section{Model}
\label{sec:Model}

To study superconductivity in a strain-induced pseudomagnetic field $\mathbf{B}_5$, we employ a
minimal model of a Weyl semimetal with a single pair of Weyl nodes separated by $2\mathbf{b}$ in
momentum space. Because of a nonzero chiral shift $\mathbf{b}$, the TR symmetry is broken in
such a model. The explicit form of the low-energy Hamiltonian reads
\begin{equation}
\label{Model-H-def}
H = \int d^3 \mathbf{x} \Psi^{\dag}(\mathbf{x}) \hat{H} \Psi(\mathbf{x}),
\end{equation}
where
\begin{eqnarray}
\label{Model-hatH}
\hat{H} &=& \left(
                    \begin{array}{cc}
                      H_{\chi=+}  & 0 \\
                      0 & H_{\chi=-} \\
                    \end{array}
                  \right),\\
\label{Model-H-chi}
H_{\chi}&=& -\mu +\chi v_F \bm{\sigma} \cdot\left(-i\bm{\nabla} +e \mathbf{A}_{\chi} -\chi \mathbf{b}\right).
\end{eqnarray}
Here $\chi=\pm$ is the chirality of Weyl nodes, $\mu$ is the electric chemical potential, $v_F$ is the Fermi velocity, $\bm{\sigma}=(\sigma_x,\sigma_y,\sigma_z)$ are the Pauli matrices, which are related to the spin or, in general, pseudospin degree of freedom,
and $\mathbf{A}_{\chi}=\mathbf{A}+\chi \mathbf{A}_5$ is the chiral vector potential, which includes
both the electromagnetic $\mathbf{A}$ and axial $\mathbf{A}_5$ vector potentials. (In the latter, there is only a coordinate-dependent part.)

In order to study the effects of pseudomagnetic fields that may result, for example, in a spatially
inhomogeneous superconducting gap, we will utilize the quasiclassical Eilenberger approach \cite{Eilenberger:1968}.
As we will discuss in the next section, such an approach uses a weak field expansion. Therefore,
the starting point in the analysis is the zeroth order solution, where the fields are absent. The structure
of possible solutions could be revealed by writing down the general form of the BdG Hamiltonian in
momentum space, i.e.,
\begin{equation}
\label{Model-HBdG}
H_{\rm BdG}(\mathbf{k}) = \left(
                    \begin{array}{cc}
                      \hat{H}(\mathbf{k}) & \hat{\Delta} \\
                      \hat{\Delta}^{\dag} & -\hat{\Theta} \hat{H}(\mathbf{k})\hat{\Theta}^{-1} \\
                    \end{array}
                  \right),
\end{equation}
where $\mathbf{k}$ is the momentum, $\hat{\Delta}$ is the gap matrix,
\begin{equation}
\label{Model-Theta-def}
\hat{\Theta} = i I_2\otimes\sigma_y \hat{K} \hat{\Pi}_{\mathbf{k}\to-\mathbf{k}}
\end{equation}
is the time-reversal operator, $I_2$ is the $2\times2$ unit matrix, $i\sigma_y $ is the Pauli matrix that describes
the spin flip, $\hat{K}$ is the complex conjugation operator, and the operator $\hat{\Pi}_{\mathbf{k}\to-\mathbf{k}}$
changes the sign of $\mathbf{k}$. The $8\times 8$ BdG Hamiltonian (\ref{Model-HBdG}) acts in the space of
the Nambu--Gor'kov spinors
\begin{equation}
\label{Model-Nambu-Psi}
\Psi_{\rm BdG}=\left\{\Psi,\Psi_{\Theta}\right\}^{T},
\end{equation}
where
\begin{equation}
\label{Model-Nambu-Psi-1}
\Psi = \left\{\psi^{\chi=+}_{\uparrow} (\mathbf{k}),\psi^{\chi=+}_{\downarrow}(\mathbf{k}),\psi^{\chi=-}_{\uparrow}(\mathbf{k}),\psi^{\chi=-}_{\downarrow}(\mathbf{k})\right\}^{T}
\end{equation}
and the TR conjugate spinor is given by
\begin{equation}
\label{Model-Nambu-Psi-2}
\Psi_{\Theta} = \left\{\psi^{\chi=+}_{\downarrow}(-\mathbf{k}),-\psi^{\chi=+}_{\uparrow}(-\mathbf{k}),\psi^{\chi=-}_{\downarrow}(-\mathbf{k}), -\psi^{\chi=-}_{\uparrow}(-\mathbf{k})\right\}^{\dag}.
\end{equation}
Here $\uparrow$ and $\downarrow$ correspond to the states with (pseudo-)spin up and down, respectively.
The structure of the gap matrix $\hat{\Delta}$ depends on the Cooper pairing channel.
In the case of inter-node pairing, which involves quasiparticles from the Weyl nodes of \emph{opposite} chirality, the corresponding
gap matrix is given by
\begin{equation}
\label{Model-Delta-BdG-Delta-inter}
\hat{\Delta}_{\rm inter} = \left(
                           \begin{array}{cc}
                             0 & \hat{\Delta}_{\chi=+} \\
                             \hat{\Delta}_{\chi=-} & 0 \\
                           \end{array}
                         \right)
 = \left(
                 \begin{array}{cc}
                   0 & \Delta_0 +\left(\bm{\Delta} \cdot \bm{\sigma}\right) \\
                   \Delta_0 -\left(\bm{\Delta} \cdot \bm{\sigma}\right) & 0 \\
                 \end{array}
               \right),
\end{equation}
and we assumed that it is independent of momentum.
As is easy to check, $\Delta_0$ corresponds to a spin-singlet state and the vector order parameter $\bm{\Delta}$ describes a spin-triplet gap. It is worth noting that a spin-triplet pairing with a momentum-independent $\bm{\Delta}$ is indeed allowed because the pairing occurs between the quasiparticles from the Weyl nodes of opposite chirality. Technically, this is related to the fact that $\left(\bm{\Delta} \cdot \bm{\sigma}\right)$ enters the antidiagonal with a different sign.

For the sake of completeness, let us also present the gap matrix for the intra-node pairing, which involves the quasiparticles from the Weyl nodes of the \emph{same} chirality, i.e.,
\begin{equation}
\label{Model-Delta-BdG-Delta-intra}
\hat{\Delta}_{\rm intra}= \left(
                           \begin{array}{cc}
                             \hat{\Delta}_{\chi=+} & 0 \\
                             0 & \hat{\Delta}_{\chi=-} \\
                           \end{array}
                         \right)
 = \left(
                 \begin{array}{cc}
                   \Delta_0 & 0 \\
                   0 & -\Delta_0  \\
                 \end{array}
               \right).
\end{equation}
Here the spin-triplet terms with a momentum-independent gap are absent because of the Pauli principle. Formally, this can be
seen from the fact that the fermion operators anticommutate $\{\psi_{\alpha}^{\chi},\psi_{\beta}^{\chi}\} =0$.

\subsection{Bogolyubov--de Gennes Hamiltonian}
\label{sec:Model-BdG-reduction-inter}

In the case of superconducting states with the inter-node pairing, the
corresponding effective BdG Hamiltonian is given by
\begin{equation}
\label{Model-BdG-reduction-HBdG-inter}
\hat{H}_{\rm BdG}(\mathbf{k}) =  \left(
                    \begin{array}{cc}
                      \hat{H}_{\chi}(\mathbf{k})  & \hat{\Delta}_{\chi} \\
                      \hat{\Delta}_{\chi}^{\dag} & i\sigma_y \hat{H}_{-\chi}^{*}(-\mathbf{k})i\sigma_y \\
                    \end{array}
                  \right) = \left(
                    \begin{array}{cc}
                      -\mu +\chi v_F\left(\bm{\sigma}\cdot\mathbf{k}_{\chi}\right)  &  \Delta_0+ \chi \left(\bm{\Delta} \cdot \bm{\sigma}\right) \\
                      \left[\Delta_0 + \chi \left(\bm{\Delta} \cdot \bm{\sigma}\right)\right]^{\dag} & \mu +\chi v_F\left(\bm{\sigma}\cdot\mathbf{k}_{\chi}\right) \\
                    \end{array}
                  \right),
\end{equation}
where $\mathbf{k}_{\chi}=\mathbf{k}-\chi\mathbf{b}$.
In order to diagonalize the kinetic part of the BdG Hamiltonian (\ref{Model-BdG-reduction-HBdG-inter}), we use the following unitary transformation:
\begin{equation}
\label{Model-BdG-reduction-HBdG-inter-diag}
\left(
  \begin{array}{cc}
    (U_{\chi})^{\dag} & 0 \\
    0 & (U_{\chi})^{\dag} \\
  \end{array}
\right)
\hat{H}_{\rm BdG}(\mathbf{k}) \left(
  \begin{array}{cc}
    U_{\chi} & 0 \\
    0 & U_{\chi} \\
  \end{array}
\right) =  \left(
    \begin{array}{cc}
        \begin{array}{cc}
           -\mu +v_Fk_{\chi} & 0 \\
          0 & -\mu -v_Fk_{\chi} \\
        \end{array}
       & (U_{\chi})^{\dag}\hat{\Delta}_{\chi} U_{\chi} \\
      (U_{\chi})^{\dag}\hat{\Delta}^{\dag}_{\chi} U_{\chi} &   \begin{array}{cc}
           \mu +v_Fk_{\chi} & 0 \\
          0 & \mu -v_Fk_{\chi} \\
        \end{array} \\
    \end{array}
  \right).
\end{equation}
Here $k_{\chi}=|\mathbf{k}_{\chi}|$ and $U_{\chi}$ is composed of the eigenvectors of $\hat{H}_{\chi}(\mathbf{k})$, i.e.,
\begin{equation}
\label{Model-BdG-reduction-U-chi-def}
U_{\chi} = \frac{1}{\sqrt{2}}\left(
             \begin{array}{cc}
               \chi e^{-i\varphi_{\chi}} \sqrt{1+\chi \cos{\theta_{\chi}}} & -\chi e^{-i\varphi_{\chi}} \sqrt{1-\chi \cos{\theta_{\chi}}} \\
               \sqrt{1-\chi \cos{\theta_{\chi}}} & \sqrt{1+\chi \cos{\theta_{\chi}}} \\
             \end{array}
           \right),
\end{equation}
where we used the spherical coordinates for the momentum: $\mathbf{k}_{\chi}=k_{\chi}\left\{\cos{\varphi_{\chi}} \sin{\theta_{\chi}}, \sin{\varphi_{\chi}} \sin{\theta}_{\chi}, \cos{\theta_{\chi}}\right\}$.
Without loss of generality, let us assume that $\mu > 0$.
The explicit form of the transformed gap reads
\begin{eqnarray}
\label{Model-BdG-reduction-Delta-chi-inter-diag}
(U_{\chi})^{\dag}\hat{\Delta}_{\chi} U_{\chi} &=& \Delta_0 +
\chi\left(\Delta_1 \cos{\varphi_{\chi}}\cos{\theta_{\chi}} +\Delta_2 \sin{\varphi_{\chi}}\cos{\theta_{\chi}} -\Delta_3 \sin{\theta_{\chi}} \right) \sigma_x
-\left(\Delta_1 \sin{\varphi_{\chi}} -\Delta_2 \cos{\varphi_{\chi}}\right) \sigma_y \nonumber\\
&+&\left(\Delta_1 \cos{\varphi_{\chi}}\sin{\theta_{\chi}} +\Delta_2 \sin{\varphi_{\chi}}\sin{\theta_{\chi}} +\Delta_3 \cos{\theta_{\chi}} \right) \sigma_z.
\end{eqnarray}
By keeping only the two dominant low-energy modes near the Fermi level (i.e., $k_{\chi}\simeq k_F\equiv \mu/v_F$), we arrive at the following reduced BdG Hamiltonian for the inter-node pairing:
\begin{equation}
\label{Model-BdG-reduction-HBdG-inter-2reduced}
\hat{H}_{\rm BdG}^{\rm (inter)}(\mathbf{k}_{\chi}) = \left(
                        \begin{array}{cc}
                          -\mu+v_F k_{\chi} & \Delta_{\chi}(\hat{\mathbf{k}}_{\chi}) \\
                          \Delta_{\chi}^{*}(\hat{\mathbf{k}}_{\chi}) & \mu-v_F k_{\chi} \\
                        \end{array}
                      \right),
\end{equation}
where the gap term is given by
\begin{equation}
\label{Model-BdG-reduction-Delta-chi-inter-def}
\Delta_{\chi}(\hat{\mathbf{k}}_{\chi}) \equiv \Delta_1 \left(\chi \cos{\varphi_{\chi}}\cos{\theta_{\chi}} +i\sin{\varphi_{\chi}}\right)
+\Delta_2 \left(\chi \sin{\varphi_{\chi}}\cos{\theta_{\chi}} -i\cos{\varphi_{\chi}}\right) -\Delta_3 \chi \sin{\theta_{\chi}}.
\end{equation}
It is worth noting that the gap function (\ref{Model-BdG-reduction-Delta-chi-inter-def}) in the reduced BdG Hamiltonian (\ref{Model-BdG-reduction-HBdG-inter-2reduced}) acquired an additional dependence on the angles of momentum $\mathbf{k_{\chi}}$. Obviously, such a dependence is induced by the unitary transformation $U_{\chi}$, which diagonalizes the kinetic part. The origin of this angular dependence is similar, e.g., to that in Ref.~\cite{Hashimoto2016}, where it appears after the transition to the band basis and projection onto the conduction band.
We checked that the contributions due to the spin singlet parameter $\Delta_0$ are suppressed at large $\mu$.
Therefore, it is not surprising that they are absent in the reduced Hamiltonian.

It is worth noting that the inter-node spin-triplet pairing always results in a gapless state \cite{Meng-Balents:2012,Cho-Moore:2012}.
For example, in the special case $\Delta_1=\Delta_2=0$,
the energy spectrum of Hamiltonian (\ref{Model-BdG-reduction-HBdG-inter-2reduced}) is given by
\begin{equation}
\label{Model-BdG-reduction-HBdG-inter-2reduced-esp-Delta3}
\epsilon_{k_{\chi}} = \pm \sqrt{(\mu-v_Fk_{\chi})^2+|\Delta_3|^2\sin^2{\theta_{\chi}}},
\end{equation}
which is indeed gapless at $\theta_{\chi}=0$. Since the inter-node pairing involves quasiparticles from the opposite chirality
Weyl nodes, it could be affected by the chiral shift. However, in the limit when the size of the gap is much smaller
than the chemical potential, the relative shift of the nodes is not very important.

In passing, let us briefly discuss the case of intra-node pairing. Performing the partial diagonalization and removing the chiral shift via the chiral transformation $\psi_{\chi}\to e^{i\chi(\mathbf{b}\cdot\mathbf{r})}\psi_{\chi}$ and
$\psi^{\dag}_{\chi}\to e^{-i\chi(\mathbf{b}\cdot\mathbf{r})}\psi^{\dag}_{\chi}$, we obtain the following reduced Hamiltonian (in general, such a transformation is anomalous and could contribute to the chiral charge density, however, this is irrelevant for the present analysis):
\begin{equation}
\label{Model-BdG-reduction-HBdG-intra-2reduced}
\hat{H}_{\rm BdG}^{\rm (intra)}(\mathbf{k}) = \left(
                        \begin{array}{cc}
                          -\mu+v_F k & \chi\Delta_0 e^{2i\chi(\mathbf{b}\cdot\mathbf{r})} \\
                          \chi\Delta_0^{*} e^{-2i\chi(\mathbf{b}\cdot\mathbf{r})} & \mu-v_F k \\
                        \end{array}
                      \right),
\end{equation}
where the additional phase factors in the gap stem from the chiral transformation.

As is easy to check, the energy spectrum of the reduced Hamiltonian (\ref{Model-BdG-reduction-HBdG-intra-2reduced}) reads
\begin{equation}
\label{Model-BdG-reduction-HBdG-intra-2reduced-esp-Delta0}
\epsilon_{k} = \pm \sqrt{(\mu-v_Fk)^2+|\Delta_0|^2}.
\end{equation}
This describes a spin-singlet state with a fully gapped spectrum. As we showed above, the same property does not hold
in the case of inter-node spin-triplet pairing because the corresponding gap vanishes at certain points in the
momentum space. These observations agree with the results in Refs.~\cite{Meng-Balents:2012,Cho-Moore:2012}.

\section{Derivation of the quasiclassical Eilenberger equation}
\label{sec:Eilenberger-Derivation}

In this section, we derive the quasiclassical Eilenberger equation for the superconducting states of Weyl materials.
Compared to the Eilenberger equation for conventional superconductors in external electromagnetic fields, see, e.g., Ref.~\cite{Houghton-Vekhter:1998}, the distinctive features of the formalism for Weyl materials will be the
relativisticlike nature of chiral quasiparticles and the presence of the constant strain-induced pseudomagnetic field $\mathbf{B}_5$.

The starting point in the derivation of the gap equations are the mean-field Gor'kov equations in the imaginary time
formalism, i.e.,
\begin{eqnarray}
\label{Model-Gor'kov-Eq-1}
\left[\partial_{\tau_1} +\hat{H}_{\rm kin} \left(\mathbf{x}_1, -i\bm{\nabla}_{\mathbf{x}_1} +\tau_z e \mathbf{A}_{\chi} (\mathbf{x}_1) \right) +\hat{\Delta}_{\rm BdG}(-i\bm{\nabla}_{\mathbf{x}_1}, \mathbf{x}_1)\right] \hat{G}_{\rm BdG}(x_1, x_2) = \delta(x_1-x_2),\\
\label{Model-Gor'kov-Eq-2}
\tau_z \hat{G}_{\rm BdG}(x_1, x_2)\left[-\partial_{\tau_2} +\hat{H}_{\rm kin} \left(\mathbf{x}_2, i\bm{\nabla}_{\mathbf{x}_2} +\tau_z e \mathbf{A}_{\chi} (\mathbf{x}_2)\right) +\hat{\Delta}_{\rm BdG}(i\bm{\nabla}_{\mathbf{x}_2}, \mathbf{x}_2)\right]\tau_z  = \delta(x_1-x_2).
\end{eqnarray}
Here $x_{1,2}=\left\{\tau_{1,2}, \mathbf{x}_{1,2}\right\}$, $\tau_{1,2}$ are imaginary times, $\mathbf{x}_{1,2}$ are spatial coordinate vectors,
$\tau_z$ is the Pauli matrix that acts in the Nambu--Gor'kov space, and $\hat{H}_{\rm kin}$ is the kinetic part of a BdG Hamiltonian, which is obtained by omitting gaps.
The general structure of Green's function is given by
\begin{equation}
\label{Model-Gor'kov-Green-def}
\hat{G}_{\rm BdG}(x_1, x_2) = \left(
                                                                   \begin{array}{cc}
                                                                     \hat{G}(x_1, x_2) & \hat{F}(x_1, x_2) \\
                                                                     \hat{F}^{\dag}(x_1, x_2) & -\hat{\bar{G}}(x_1, x_2) \\
                                                                   \end{array}
                                                                 \right),
\end{equation}
where
\begin{eqnarray}
\label{Model-Gor'kov-Green-G-def}
\hat{G}_{\rm BdG}(x_1,x_2) &=& \left\langle T_{\tau} \Psi(x_1) \Psi^{\dag}(x_2) \right\rangle,\\
\label{Model-Gor'kov-Green-Gbar-def}
\hat{\bar{G}}_{\rm BdG}(x_1,x_2) &=& -\left\langle T_{\tau} \Psi^{\dag}(x_1) \Psi(x_2) \right\rangle
\end{eqnarray}
are particle and hole propagators and $T_{\tau}$ is ordering operator in the imaginary time.
By definition, the anomalous Gor'kov functions are
\begin{eqnarray}
\label{Model-Gor'kov-Green-F-def}
\hat{F}_{\rm BdG}(x_1,x_2) &=& \left\langle T_{\tau} \Psi(x_1) \Psi(x_2) \right\rangle,\\
\label{Model-Gor'kov-Green-Fdag-def}
\hat{F}^{\dag}_{\rm BdG}(x_1,x_2) &=& \left\langle T_{\tau} \Psi^{\dag}(x_1) \Psi^{\dag}(x_2) \right\rangle .
\end{eqnarray}
In general, the Gor'kov equations (\ref{Model-Gor'kov-Eq-1}) and (\ref{Model-Gor'kov-Eq-2}) should also include
the self-energy terms. In this study, for simplicity, we neglect the corresponding effects.

Since we do not consider time-dependent background fields, it is convenient to
work with the Fourier transforms of Green's function in the imaginary time, i.e.,
\begin{equation}
\label{Model-Gor'kov-Green-Matsubara}
\hat{G}_{\rm BdG}(x_1, x_2) =i T\sum_{m^{\prime}=-\infty}^{\infty} \hat{G}_{\rm BdG}(\mathbf{x}_1, \mathbf{x}_2;i\omega_{m^{\prime}})
e^{-i\omega_{m^{\prime}}(\tau_1-\tau_2)},
\end{equation}
where $\omega_{m^{\prime}}=\pi T (2m^{\prime}+1)$ are the fermionic Matsubara frequencies and $T$ is temperature.
By following Ref.~\cite{Houghton-Vekhter:1998}, it is also convenient to perform the following transformations:
\begin{eqnarray}
\label{Model-Gor'kov-new-vars-H}
\tilde{H}&=&\hat{H}_{\rm kin}\tau_z,\\
\label{Model-Gor'kov-new-vars-Delta}
\tilde{\Delta}&=&\hat{\Delta}_{\rm BdG}\tau_z,\\
\label{Model-Gor'kov-new-vars-G}
\tilde{G}&=&\tau_z\hat{G}_{\rm BdG}.
\end{eqnarray}
Then, Eqs.~(\ref{Model-Gor'kov-Eq-1}) and (\ref{Model-Gor'kov-Eq-2}) take the form
\begin{eqnarray}
\label{Model-Gor'kov-Eq-New-1}
&&\left[-i\omega_{m^{\prime}} \tau_z +\tilde{H} \left(\mathbf{x}_1, -i\bm{\nabla}_{\mathbf{x}_1}
+\tau_z e \mathbf{A}_{\chi} (\mathbf{x}_1)\right) +\tilde{\Delta}( -i\bm{\nabla}_{\mathbf{x}_1},\mathbf{x}_1)\right]
\tilde{G}(\mathbf{x}_1, \mathbf{x}_2; i\omega_{m^{\prime}}) = \delta(\mathbf{x}_1-\mathbf{x}_2),\\
\label{Model-Gor'kov-Eq-New-2}
&&\tilde{G}(\mathbf{x}_1, \mathbf{x}_2; i\omega_{m^{\prime}})\left[-i\omega_{m^{\prime}} \tau_z
+\tilde{H} \left(\mathbf{x}_2, i\bm{\nabla}_{\mathbf{x}_2} +\tau_z e \mathbf{A}_{\chi} (\mathbf{x}_2)\right)
+\tilde{\Delta}(i\bm{\nabla}_{\mathbf{x}_2},\mathbf{x}_2)\right]  = \delta(\mathbf{x}_1-\mathbf{x}_2).
\end{eqnarray}
While the quasiclassical Green's functions are not translationally invariant in the presence of weak and slowly varying
background fields, their dependence on the center-of-mass coordinate $\mathbf{R} =(\mathbf{x}_1+\mathbf{x}_2)/2$
should be weak. In this case, it is reasonable to use a gradient expansion. In order to obtain such a systematic expansion,
it is convenient to rewrite the Green's functions in terms of the center-of-mass coordinate $\mathbf{R}$ and the relative
coordinate $\mathbf{r}=\mathbf{x}_1-\mathbf{x}_2$. Furthermore, we perform the Fourier transform
with respect to the relative coordinate $\mathbf{r}$, i.e.,
\begin{equation}
\label{Model-Gor'kov-Green-integrated}
\tilde{G}(\mathbf{k}, \mathbf{R};i\omega_{m^{\prime}}) = \int d^3\mathbf{r} e^{-i\mathbf{k}\mathbf{r}}
\tilde{G}\left(\mathbf{R}+\frac{\mathbf{r}}{2},\mathbf{R}-\frac{\mathbf{r}}{2};i\omega_{m^{\prime}}\right).
\end{equation}
By making use of this Green's function, the Gor'kov equations (\ref{Model-Gor'kov-Eq-New-1}) and (\ref{Model-Gor'kov-Eq-New-2})
can be rewritten as follows \cite{Houghton-Vekhter:1998}:
\begin{eqnarray}
\label{Model-Gor'kov-Eq-New-FT-1}
&&\left[-i\omega_{m^{\prime}} \tau_z +\tilde{H} \left(\mathbf{k} +\tau_z e \mathbf{A}_{\chi}(\mathbf{R})\right) +\tilde{\Delta}(\mathbf{k}, \mathbf{R})\right] \circ \tilde{G}(\mathbf{k}, \mathbf{R}; i\omega_{m^{\prime}}) = 1,\\
\label{Model-Gor'kov-Eq-New-FT-2}
&&\tilde{G}(\mathbf{k}, \mathbf{R}; i\omega_{m^{\prime}}) \circ \left[-i\omega_{m^{\prime}} \tau_z +\tilde{H} \left(\mathbf{k} +\tau_z e \mathbf{A}_{\chi}(\mathbf{R})\right) +\tilde{\Delta}(\mathbf{k},\mathbf{R})\right]  = 1,
\end{eqnarray}
where we employed the circle product \cite{Eckern-Schmid:1981,Houghton-Vekhter:1998}, which is formally defined for some functions $A(\mathbf{k}, \mathbf{R})$ and $B(\mathbf{k}, \mathbf{R})$ as
\begin{equation}
\label{Model-Gor'kov-circle-def}
A(\mathbf{k}, \mathbf{R}) \circ B(\mathbf{k}, \mathbf{R}) = \lim_{\mathbf{R}_{1,2}\to\mathbf{R}}
\lim_{\mathbf{k}_{1,2}\to\mathbf{k}} e^{\frac{i}{2}\left(\bm{\nabla}_{\mathbf{k}_2}\bm{\nabla}_{\mathbf{R}_1}
- \bm{\nabla}_{\mathbf{k}_1}\bm{\nabla}_{\mathbf{R}_2}\right)} A(\mathbf{k}_1, \mathbf{R}_1) B(\mathbf{k}_2, \mathbf{R}_2).
\end{equation}
The form of the Gor'kov equations (\ref{Model-Gor'kov-Eq-New-FT-1}) and (\ref{Model-Gor'kov-Eq-New-FT-2})
is well suited for a systematic expansion in powers of background fields and spatial gradients with respect to the
center-of-mass coordinate. As we will see in Sec.~\ref{sec:Eilenberger-Sol-inter-Gap}, in order to study the
effects of the pseudomagnetic field on the gap generation, the expansion must be performed up to the second order.
The corresponding explicit expansion of Eq.~(\ref{Model-Gor'kov-Eq-New-FT-1}) is given in Appendix~\ref{sec:App-expansion} by Eq.~(\ref{Model-Gor'kov-Eq-FT-1-all}).

The final step in the derivation of the quasiclassical Eilenberger equations is to remove the dependence on the
components of momenta perpendicular to the Fermi surface. This can be achieved by integrating
over the quasiparticle energy \cite{Eilenberger:1968,Houghton-Vekhter:1998}, i.e.,
\begin{equation}
\label{Eilenberger-Eq-Green-int}
\tilde{g}(\mathbf{R}, \mathbf{k}_{\parallel}; i\omega_{m^{\prime}}) = \left(
                                                                        \begin{array}{cc}
                                                                          g(\mathbf{R}, \mathbf{k}_{\parallel}; i\omega_{m^{\prime}}) & f(\mathbf{R}, \mathbf{k}_{\parallel}; i\omega_{m^{\prime}}) \\
                                                                          -f^{\dag}(\mathbf{R}, \mathbf{k}_{\parallel}; i\omega_{m^{\prime}}) & \bar{g}(\mathbf{R}, \mathbf{k}_{\parallel}; i\omega_{m^{\prime}}) \\
                                                                        \end{array}
                                                                      \right)
= \frac{v_F}{i\pi} \int d k \tilde{G}(\mathbf{R}, \mathbf{k}; i\omega_{m^{\prime}}).
\end{equation}
In addition to its weak dependence on $\mathbf{R}$, this integrated Green's function depends only on the components
of the momentum $\mathbf{k}_{\parallel}$ parallel to the Fermi surface. As should be clear, the use of $\tilde{g}$ instead
of $\tilde{G}$ makes sense only when the electric chemical potential is significantly larger that the value of superconducting
gap. Indeed, in such a case, while the nonintegrated Green's function $\tilde{G}(\mathbf{R}, \mathbf{k}; i\omega_{m^{\prime}})$
may vary rapidly with momentum near the Fermi level, its integrated counterpart $\tilde{g}(\mathbf{R}, \mathbf{k}_{\parallel};
i\omega_{m^{\prime}})$ depends weakly on $\mathbf{k}_{\parallel}$ \cite{Eilenberger:1968,Houghton-Vekhter:1998}. Integrating over the quasiparticle energy in Eq.~(\ref{Model-Gor'kov-Eilenberger-Eq-diff}), we obtain the following Eilenberger equation:
\begin{eqnarray}
\label{Model-Gor'kov-Eilenberger-Eq-final}
&&-i\omega_{m^{\prime}} \left[\tau_z, \tilde{g}\right]
+ eA_{\chi}^{j}\left[\tau_z \tilde{V}_j, \tilde{g}\right]
-\frac{i}{2} \left\{\tilde{V}_j, (\partial_{R_{j}}\tilde{g})\right\}
+ \frac{i e}{2} \epsilon^{jlm}B_{\chi}^{m}
\left\{\tau_z \tilde{V}_l, (\partial_{k_{\parallel,j}}\tilde{g})\right\}
+\frac{e^2}{2}A^{j}_{\chi} A^{l}_{\chi}\left[\tilde{M}_{jl}, \tilde{g}\right] \nonumber\\
&&+\frac{ie^2}{2} A^{l}_{\chi} \epsilon^{j m n}B_{\chi}^{n}
\left\{\tilde{M}_{ml}, (\partial_{k_{\parallel,j}} \tilde{g}) \right\}
-\frac{ie}{2} A^{l}_{\chi} \left\{\tau_z \tilde{M}_{jl}, (\partial_{R_{j}} \tilde{g}) \right\}
-\frac{e^2}{8} \epsilon^{l m p} \epsilon^{j n s} B_{\chi}^{p} B_{\chi}^{s} \left[\tilde{M}_{mn}, (\partial_{k_{\parallel, j}}\partial_{k_{\parallel, l}} \tilde{g})\right]
\nonumber\\
&&
+\frac{e}{4} \epsilon^{l m n} B_{\chi}^{n} \left[\tau_z\tilde{M}_{mj}, (\partial_{k_{\parallel, l}}\partial_{R_{j}} \tilde{g})\right]
-\frac{1}{8} \left[\tilde{M}_{jl}, (\partial_{R_{j}}\partial_{R_{l}} \tilde{g}) \right] +\left[\tilde{\Delta}, \tilde{g}\right]
+\frac{i}{2} \left\{(\partial_{R_{j}} \tilde{\Delta}), (\partial_{k_{\parallel,j}} \tilde{g}) \right\}
\nonumber\\
&&
-\frac{i}{2} \left\{(\partial_{k_{\parallel, j}}\tilde{\Delta}), (\partial_{R_{j}}\tilde{g}) \right\}
-\frac{1}{8} \left[(\partial_{R_{j}}\partial_{R_{l}} \tilde{\Delta}), (\partial_{k_{\parallel, j}}\partial_{k_{\parallel, l}}\tilde{g}) \right]
+\frac{1}{4} \left[(\partial_{k_{\parallel, j}}\partial_{R_{l}}\tilde{\Delta}), (\partial_{R_{j}}\partial_{k_{\parallel, l}}\tilde{g})\right]
\nonumber\\
&&
-\frac{1}{8} \left[(\partial_{k_{\parallel, j}}\partial_{k_{\parallel, l}} \tilde{\Delta}), (\partial_{R_{j}}\partial_{R_{l}}\tilde{g})\right]
+O\left(\partial_{R_{j}}^3, (A_{\chi}^{j})^3,(\partial_{R_j} A_{\chi}^{l})^3\right)=0.
\end{eqnarray}
Here, for the sake of brevity,
we used the Einstein summation convention and omitted the arguments $\mathbf{k}_{\parallel}$,
$\hat{\mathbf{k}}$, and $\mathbf{R}$ in functions $\tilde{g}$, $\tilde{H}$, and $\tilde{\Delta}$. By definition, $\epsilon^{jlm}$ is the
antisymmetric Levi-Civita tensor and the square (curly) brackets denote the commutators (anticommutators).
We also took into account that the kinetic part of the BdG Hamiltonian $\tilde{H}$ is proportional to the unit matrix in the
Nambu--Gor'kov space and used the following shorthand notations:
\begin{eqnarray}
\label{Model-Gor'kov-Eilenberger-V-def}
\tilde{\mathbf{V}} &\equiv& (\bm{\nabla}_{\mathbf{k}}\tilde{H}),\\
\label{Model-Gor'kov-Eilenberger-M-def}
\tilde{M}_{jl} &\equiv&(\partial_{k_{j}} \partial_{k_{l}}\tilde{H}).
\end{eqnarray}
In general, the Eilenberger equation (\ref{Model-Gor'kov-Eilenberger-Eq-final}) is reminiscent of a kinetic equation, where
the role of a distribution function is played by the integrated Green's function $\tilde{g}$. It is instructive to comment briefly
on the physical meaning of some terms in Eq.~(\ref{Model-Gor'kov-Eilenberger-Eq-final}).
For example, the covariant derivative $-i\partial_{R_{j}}+2eA^{j}_{\chi}$, where the doubled electric charge $2e$ is indeed expected for Cooper pairs, is related to the second and third terms. The (pseudo-)Lorentz force is described by the fourth term.
As we see, there are also several terms related to the spatial and momentum dependence of a gap.

It should be noted that, unlike the original Gor'kov equations, Eq.~(\ref{Model-Gor'kov-Eilenberger-Eq-final}) is homogeneous and, therefore, it is not sufficient by itself to determine unambiguously function $\tilde{g}$.
As argued in Refs.~\cite{Eilenberger:1968,Larkin-Ovchinnikov:1968-g2,Houghton-Vekhter:1998,Kopnin-book:2001,Matsushita:2018bty}, it should be supplemented by the following normalization condition:
\begin{equation}
\label{Model-Gor'kov-Eilenberger-g-norm-Eq}
\tilde{g}^2 = 1.
\end{equation}
Such a condition holds when the self-energy terms in the Gor'kov equations and, consequently, in the Eilenberger
equation are omitted \cite{Kopnin-book:2001,Matsushita:2018bty}.

\section{Solutions to the Eilenberger equation}
\label{sec:Eilenberger-Eqs-sol}

In this section, we determine the superconducting gaps and calculate electric and chiral currents in strained Weyl semimetals by using the iterative solutions of the Eilenberger equation (\ref{Model-Gor'kov-Eilenberger-Eq-final}) amended by the normalization condition (\ref{Model-Gor'kov-Eilenberger-g-norm-Eq}).
Before proceeding to the calculations, it is instructive to qualitatively
discuss the difference between the effects of the pseudomagnetic field $\mathbf{B}_5$ on the intra- and inter-node
pairings in Weyl semimetals. In the former case, the pairing occurs between quasiparticles from the same Weyl node
and, as a result, the chiral charge of Cooper pairs is twice as large as that of the individual quasiparticles. Since
the coupling of the pseudomagnetic field to quasiparticles is proportional to their chirality, its effect on the intra-node
Cooper pairs from each of Weyl nodes is similar to that of a usual magnetic field. In such a case, the superconducting
currents and gaps become spatially nonuniform. This complicates significantly the study of the corresponding phase, which
will be only briefly discussed at the end of Sec.~\ref{sec:Eilenberger-Sol-inter-Gap} in the limit of the vanishing fields.
It should be also noted that the source of such a nontrivial spatial dependence is the axial vector potential $\mathbf{A}_5$,
which, unlike its electromagnetic counterpart, is an observable quantity by itself.

In the case of inter-node pairing, on the other hand, the Cooper pairs are made from the quasiparticles
of opposite chirality that couple to the pseudomagnetic field with opposite signs. Therefore, the
corresponding phase has a simpler structure. This is one of the main reasons why our analysis in the present paper will be
concentrated primarily on the superconducting states with the inter-node pairing.
Also, we will assume that the ordinary magnetic field is absent, i.e., $\mathbf{A}_{\chi}=\chi\mathbf{A}_5$ and $\mathbf{B}_{\chi}=\chi\mathbf{B}_5$.

In order to solve the Eilenberger equation (\ref{Model-Gor'kov-Eilenberger-Eq-final}), the iterative method is used. In particular, we expand the integrated Green's function $\tilde{g}$ up to the second order in powers of the background fields and spatial derivatives, i.e.,
\begin{equation}
\label{Eilenberger-sol-inter-g-expand}
\tilde{g}\approx \tilde{g}_0+\tilde{g}_1+\tilde{g}_2,
\end{equation}
where the subscript corresponds to the order of expansion.
The details of the derivation, including the explicit expressions for the integrated Green's function components in Eq.~(\ref{Eilenberger-sol-inter-g-expand}), are given in Appendix \ref{sec:Eilenberger-sol-inter}.

\subsection{Superconducting gaps}
\label{sec:Eilenberger-Sol-inter-Gap}

In this subsection, we analyze the effects of the strain-induced pseudomagnetic field $\mathbf{B}_5$ on the
generation of a superconducting gap in Weyl semimetals. Without loss of generality, we assume that
$\mathbf{B}_5\parallel \hat{\mathbf{z}}$. The general form of the gap equation reads
\begin{equation}
\label{Eilenberger-Sol-inter-Gap-Delta-Eq}
\Delta_{\chi}(\mathbf{k}_{\parallel},\mathbf{R}) = \rho(\mu) \pi iT \sum_{m^{\prime}=-\infty}^{\infty} \int \frac{d\Omega_{\mathbf{k}^{\prime}_{\parallel}}}{4\pi} U\left(\mathbf{k}_{\parallel},\mathbf{k}^{\prime}_{\parallel}\right) f(\mathbf{k}^{\prime}_{\parallel}, \mathbf{R}),
\end{equation}
where $\rho(\mu)$ is the density of states at the Fermi level, $\sum_{m^{\prime}=-\infty}^{\infty}$ is the
summation over the Matsubara frequencies, $\int d\Omega_{\mathbf{k}^{\prime}_{\parallel}}$ denotes
the integration over the Fermi surface, and $U\left(\mathbf{k}_{\parallel},\mathbf{k}^{\prime}_{\parallel}\right)$
is an attractive interaction potential. The anomalous part of the integrated Green's function
$f(\mathbf{k}^{\prime}_{\parallel}, \mathbf{R}) \approx f_0(\mathbf{k}^{\prime}_{\parallel}, \mathbf{R}) +f_1(\mathbf{k}^{\prime}_{\parallel}, \mathbf{R}) +f_2(\mathbf{k}^{\prime}_{\parallel}, \mathbf{R})$ is given by
Eqs.~(\ref{Eilenberger-Sol-inter-Zero-sol-f0}), (\ref{Eilenberger-Sol-inter-First-sol-f1}), and (\ref{Eilenberger-Sol-inter-Second-sol-f2}) in Appendix \ref{sec:Eilenberger-sol-inter}.
Integrating over the angles first, we can safely omit the subscript $\chi$ in the corresponding
momenta.

In the case of inter-node pairing, the structure of the gap function ansatz follows directly from Eq.~(\ref{Model-BdG-reduction-Delta-chi-inter-def}) and is given by
\begin{equation}
\label{Eilenberger-Sol-inter-Gap-Delta-chi}
\Delta_{\chi} =|\Delta_1| e^{i\mathbf{Q}_1\mathbf{R}} \left(\chi \cos{\varphi}\cos{\theta} +i\sin{\varphi}\right) +|\Delta_2|e^{i\mathbf{Q}_2\mathbf{R}} \left(\chi \sin{\varphi}\cos{\theta} -i\cos{\varphi}\right) -|\Delta_3|e^{i\mathbf{Q}_3\mathbf{R}} \chi \sin{\theta},
\end{equation}
where $|\Delta_1|$, $|\Delta_2|$, and $|\Delta_3|$ quantify the spin-triplet state with a spin projection on the axes $\hat{\mathbf{x}}$, $\hat{\mathbf{y}}$, and $\hat{\mathbf{z}}$, respectively.
The spatial modulation is described by the constant phase vectors $\mathbf{Q}_1$, $\mathbf{Q}_2$, and $\mathbf{Q}_3$.
Further, we will consider each of the spin-triplet channels separately. It is worth reminding that the angular dependence in Eq.~(\ref{Eilenberger-Sol-inter-Gap-Delta-chi}) appeared due to the reduction of the effective BdG Hamiltonian discussed in Sec.~\ref{sec:Model-BdG-reduction-inter}.
In order to be consistent with the gap equation (\ref{Eilenberger-Sol-inter-Gap-Delta-Eq}), the interaction potential $U\left(\mathbf{k}_{\parallel},\mathbf{k}^{\prime}_{\parallel}\right)=U_j\left(\mathbf{k}_{\parallel},\mathbf{k}^{\prime}_{\parallel}\right)$, where $j=1,2,3$ correspond to spin-triplet states, is chosen in the following form:
\begin{eqnarray}
\label{Eilenberger-Sol-inter-Gap-V-1}
U_1\left(\mathbf{k}_{\parallel},\mathbf{k}^{\prime}_{\parallel}\right) &=& |U| \left(\chi \cos{\varphi}\cos{\theta} +i\sin{\varphi}\right) \left(\chi \cos{\varphi^{\prime}}\cos{\theta^{\prime}} -i\sin{\varphi^{\prime}}\right),\\
\label{Eilenberger-Sol-inter-Gap-V-2}
U_2\left(\mathbf{k}_{\parallel},\mathbf{k}^{\prime}_{\parallel}\right) &=& |U| \left(\chi \sin{\varphi}\cos{\theta} -i\cos{\varphi}\right) \left(\chi \sin{\varphi^{\prime}}\cos{\theta^{\prime}} +i\cos{\varphi^{\prime}}\right),\\
\label{Eilenberger-Sol-inter-Gap-V-3}
U_3\left(\mathbf{k}_{\parallel},\mathbf{k}^{\prime}_{\parallel}\right) &=& |U| \sin{\theta} \sin{\theta^{\prime}}.
\end{eqnarray}
Here $|U|$ is the strength of the potential. It is worth noting that the form of the interaction potential is very important
for the competition between different types of Cooper pairing. In our study, however, we do not attempt to rigorously address such
a competition, but concentrate primarily on the effects of strain-induced pseudomagnetic field on the inter-node
spin-triplet pairing (i.e., the pairing between the quasiparticles from different Weyl nodes where the Cooper pairs have nonzero spin projections).
For our purposes, it is sufficient to use the fact that any interaction potential can be decomposed into spherical
harmonics with specific weights in each angular momentum channel (see, e.g., Ref.~\cite{Mineev-book:1998}).
Therefore, by assuming that the appropriate interaction channel exists, we define the interaction potentials as
$U_j\left(\mathbf{k}_{\parallel},\mathbf{k}^{\prime}_{\parallel}\right)$ in Eqs.~(\ref{Eilenberger-Sol-inter-Gap-V-1})--(\ref{Eilenberger-Sol-inter-Gap-V-3}), whose angular dependence is dictated by the self-consistency of the gap equation (\ref{Eilenberger-Sol-inter-Gap-Delta-Eq}) with the gap function (\ref{Eilenberger-Sol-inter-Gap-Delta-chi}).

By using the explicit form of $f_0$, $f_1$, and $f_2$, presented in Appendix \ref{sec:Eilenberger-sol-inter} by Eqs.~(\ref{Eilenberger-Sol-inter-Zero-sol-f0}), (\ref{Eilenberger-Sol-inter-First-sol-f1}), and
(\ref{Eilenberger-Sol-inter-Second-sol-f2}), respectively, expanding in small $|\Delta_1|/T$, integrating over the angles, and performing the
summation over the Matsubara frequencies, we obtain the following Ginzburg--Landau equation for $\Delta_1$:
\begin{eqnarray}
\label{Eilenberger-Sol-inter-Gap-Delta1-all-Q}
&&\Delta_1 \ln{\left(\frac{\omega_{\rm D}}{\pi T_0}\right)}-\Delta_1 \ln{\left(\frac{\omega_{\rm D}}{\pi T}\right)} \approx
\Delta_1 \frac{T-T_0}{T_0}
=
-\frac{\Delta_1 |\Delta_1|^2}{\pi^2 T^2} \frac{7 \zeta(3)}{10} \nonumber\\
&&-\frac{\Delta_1 v_F^2}{1120\pi^4T^4 \mu^2}
\Bigg\{
2Q_{1,x}^2 \left[49 \pi^2T^2 \mu^2 \zeta(3) -|\Delta_1|^2 \left(56\pi^2T^2\zeta(3)+93\mu^2\zeta(5)\right) \right] \nonumber\\
&&+(Q_{1,y}^2+Q_{1,z}^2) \left[196 \pi^2T^2 \mu^2 \zeta(3) +|\Delta_1|^2 \left(7\pi^2T^2\zeta(3)-558\mu^2\zeta(5)\right) \right] \nonumber\\
&&+14 \chi v_FQ_{1,z} eB_5 \mu \left[35\pi^2T^2 \zeta(3)-93|\Delta_1|^2 \zeta(5)\right]
+35 v_F^2 e^2 B_5^2 \left[14\pi^2T^2\zeta(3)-31|\Delta_1|^2\zeta(5)\right]
\Bigg\} +O(|\Delta_1|^4).
\end{eqnarray}
Since the above equation was obtained by expanding in powers of $|\Delta_1|$, it is valid only for a sufficiently small gap in the vicinity of the superconducting transition temperature $T_0$. It is worth noting that we followed the standard approach in obtaining the Ginzburg--Landau equation (see, e.g., Ref.~\cite{Landau:t9}), where the near-critical regime $(T-T_0)/T_0\ll1$ is assumed. Such a regime allowed us to perform the integration over the angles analytically. For $B_5=0$ and $\mathbf{Q}_1=\mathbf{0}$, the corresponding critical temperature $T_0$ is given by
\begin{equation}
\label{Eilenberger-Sol-inter-Gap-Delta1-Tc-def}
T_0=\frac{\omega_{\rm D}}{\pi}\exp\left(-\frac{3}{2|U|\rho(\mu)}\right).
\end{equation}
Here $\omega_{\rm D}$ is the Debye frequency, which defines the energy cutoff.
In the derivation of Eq.~(\ref{Eilenberger-Sol-inter-Gap-Delta1-all-Q}), we regularized the sum over Matsubara frequencies as
\begin{equation}
\label{Eilenberger-Sol-inter-Gap-Delta1-zeta-0-def}
\sum_{m^{\prime}=-\infty}^{\infty} \frac{1}{2m^{\prime}+1} \simeq \ln{\left(\frac{\omega_{\rm D}}{\pi T}\right)}.
\end{equation}
We also employed the standard summation formula
\begin{equation}
\label{Eilenberger-Sol-inter-Gap-Delta1-zeta-def}
\sum_{m^{\prime}=-\infty}^{\infty} \frac{1}{|2m^{\prime}+1|^{x}} = 2\left(1-\frac{1}{2^x}\right) \zeta(x),
\end{equation}
which is valid at $x>1$ and where $\zeta(x)$ is the zeta function. We note that the result of summation over $m^{\prime}$
vanishes for any function odd in $\omega_{m^{\prime}}$ and, thus, there is no contribution to the gap equation from the first order anomalous function $f_1$, which is explicitly defined in Appendix \ref{sec:Eilenberger-sol-inter} by Eq.~(\ref{Eilenberger-Sol-inter-First-sol-f1}).

The gap equations for $\Delta_{2}$ and $\Delta_{3}$ have the same form as
Eq.~(\ref{Eilenberger-Sol-inter-Gap-Delta1-all-Q}) but with the replacements
$\mathbf{Q}_1\to\mathbf{Q}_2$ and $\mathbf{Q}_1\to\mathbf{Q}_3$, respectively.
Additionally, one should interchange $Q_{2,x} \leftrightarrow Q_{2,y}$ and $Q_{3,x}
\leftrightarrow Q_{3,z}$. Last but not least, the pseudomagnetic field strength does not enter
the gap equation when the spins of Cooper pairs are parallel to the field.
Technically, this means that, in addition to the above redefinitions, one should also set $B_5\to0$, when obtaining the gap equation for $\Delta_{3}$ from Eq.~(\ref{Eilenberger-Sol-inter-Gap-Delta1-all-Q}).

\begin{figure}[t]
\begin{center}
\includegraphics[width=0.45\textwidth]{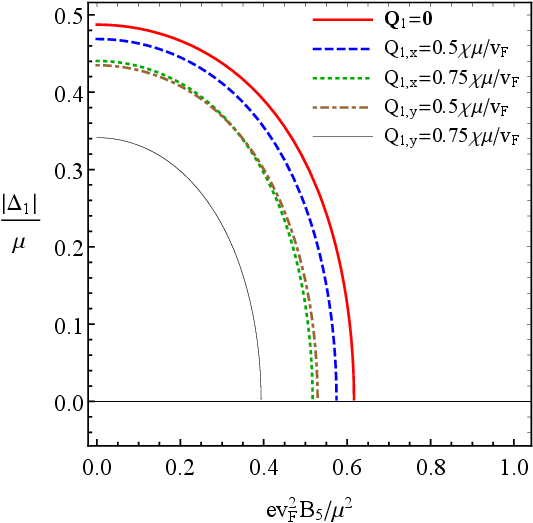}\hfill
\includegraphics[width=0.45\textwidth]{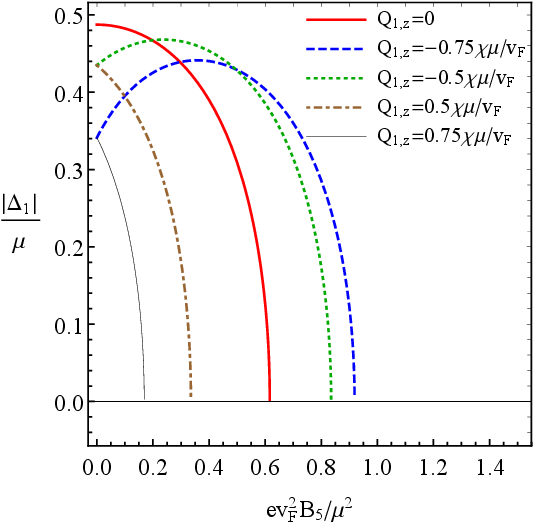}
\caption{The dependence of the absolute value of gap $|\Delta_1|$ given in
Eq.~(\ref{Eilenberger-Sol-inter-Gap-Delta1-Qnot0}) on the pseudomagnetic field strength $B_5$ at a few fixed
values of the phase vector components $Q_{1,x}$ and $Q_{1,y}$ (left panel), and $Q_{1,z}$ (right panel).
While the pseudomagnetic field tends to inhibit the gap, this effect can be mitigated when $\mathbf{Q}$ is antiparallel to $\mathbf{B}_5$.
To plot the results we used $T=0.45\mu$ and $T_0=0.5\mu$.}
\label{fig:Eilenberger-Sol-inter-Gap-Delta1-B5}
\end{center}
\end{figure}

\begin{figure}[t]
\begin{center}
\hspace{-0.3\textwidth}(a)\hspace{0.37\textwidth}(b)\hspace{0.3\textwidth}(c)\\[0pt]
\includegraphics[width=0.3\textwidth]{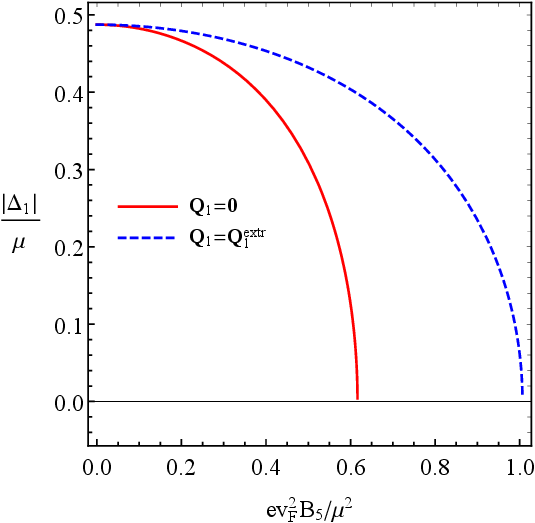}\hfill
\includegraphics[width=0.37\textwidth]{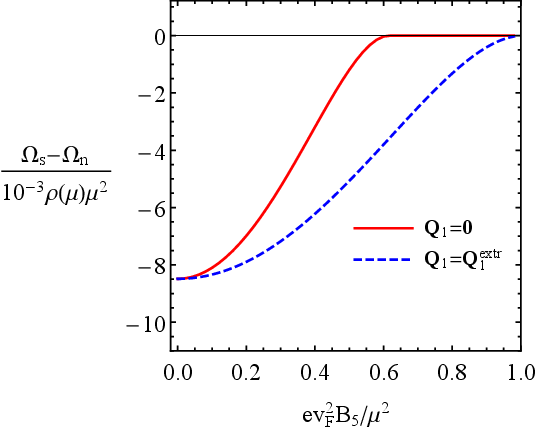}\hfill
\includegraphics[width=0.3\textwidth]{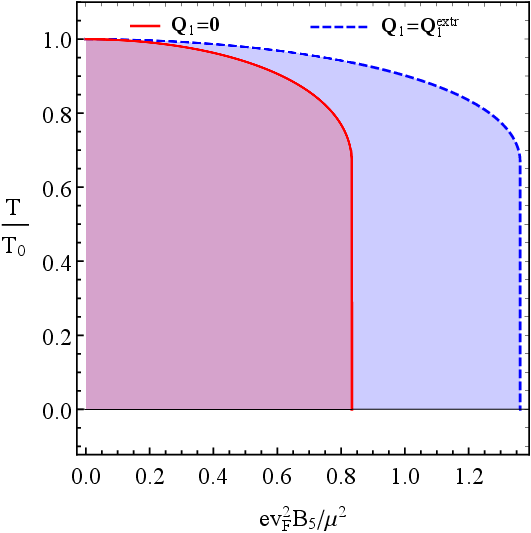}
\caption{The absolute value of gap $|\Delta_1|$ (panel (a)), the difference of the energy densities $\Omega_{\rm s} -\Omega_{\rm n}$ (panel (b)), and the phase diagram in $T$--$B_5$ plane (panel (c)) in the absence of the phase $\mathbf{Q}_1=\mathbf{0}$ (red solid lines) and for the extremal value of the phase vector $\mathbf{Q}_1=\mathbf{Q}_{1}^{\rm extr}$ (blue dashed lines), where $\mathbf{Q}_{1}^{\rm extr}$ is given by Eqs.~(\ref{Eilenberger-Sol-inter-Gap-Delta1-Q1-extr-xy}) and (\ref{Eilenberger-Sol-inter-Gap-Delta1-Q1-extr-z}).
The case with a nonzero phase vector $\mathbf{Q}_1=\mathbf{Q}_{1}^{\rm extr}$ corresponds to a more energetically favourable solution with a larger gap and critical temperature.
The shaded areas in panel (c) correspond to superconducting phases. To plot the results we used $T_0=0.5\mu$ and $T=0.45\mu$ in panels (a) and (b).}
\label{fig:Eilenberger-Sol-inter-Gap-Delta1-Omega-B5-Extr}
\end{center}
\end{figure}

The explicit form of the nontrivial solution to the gap equation (\ref{Eilenberger-Sol-inter-Gap-Delta1-all-Q})
is given by
\begin{eqnarray}
\label{Eilenberger-Sol-inter-Gap-Delta1-Qnot0}
|\Delta_1| &=& \frac{\sqrt{14}\pi T}{\sqrt{T_0}} \Bigg\{80\pi^2T^2(T_0-T)\mu^2 -7v_F^2T_0\zeta(3) \left[\left(Q_{1,x}^2+2Q_{1,y}^2+2Q_{1,z}^2\right)\mu^2 +5v_FeB_5 \left(v_FeB_5 +\chi \mu Q_{1,z}\right) \right] \Bigg\}^{1/2} \nonumber\\
&\times&\Bigg\{7\zeta(3)\pi^2 T^2\left[112\mu^2 -v_F^2\left(16Q_{1,x}^2-Q_{1,y}^2-Q_{1,z}^2\right)\right] -31\zeta(5)v_F^2\Big[6\mu^2\left(Q_{1,x}^2+3Q_{1,y}^2+3Q_{1,z}^2\right) \nonumber\\
&+&7v_FeB_5\left(5v_FeB_5+6\chi \mu Q_{1,z}\right)\Big] \Bigg\}^{-1/2}.
\end{eqnarray}
Note that when both $B_5=0$ and $\mathbf{Q}_1=\mathbf{0}$, we obtain the standard Ginzburg--Landau
value for the gap, i.e.,
\begin{equation}
\label{Eilenberger-Sol-inter-Gap-Delta1-B5=0-Q=0}
\lim_{B_5\to0}\lim_{\mathbf{Q}_1\to\mathbf{0}}|\Delta_1| = \pi T \sqrt{\frac{10}{7\zeta(3)}} \sqrt{\frac{T_0-T}{T_0}}.
\end{equation}
Further, we can also determine the critical value of the pseudomagnetic field from the condition of vanishing gap.
The corresponding expression reads
\begin{equation}
\label{Eilenberger-Sol-inter-Gap-Delta1-B5-crit}
B_5^{\rm crit} = -\frac{\chi Q_{1,z} \mu}{2 ev_F} + \frac{\mu}{2\sqrt{35 \zeta(3)T_0} ev_F^2} \sqrt{320\pi^2 T^2(T_0-T) -7\zeta(3)v_F^2T_0 \left(4Q_{1,x}^2+8Q_{1,y}^2+3Q_{1,z}^2\right)}.
\end{equation}
[Note that the above expression is valid in the vicinity of the critical temperature, which, as we will see from Fig.~\ref{fig:Eilenberger-Sol-inter-Gap-Delta1-Omega-B5-Extr}(c), is lower than $T_0$ for $B_5\neq0$ and $Q_1\neq0$.]
The dependence of the absolute value of gap $|\Delta_1|$ on the pseudomagnetic field strength is presented
in Fig.~\ref{fig:Eilenberger-Sol-inter-Gap-Delta1-B5} for several fixed values of $Q_{1,x}$, $Q_{1,y}$, and
$Q_{1,z}$. As we see from the left panel in Fig.~\ref{fig:Eilenberger-Sol-inter-Gap-Delta1-B5}, nonzero
phases $Q_{1,x}$ and $Q_{1,y}$ always reduce the gap. In agreement with the analytical expression
in Eq.~(\ref{Eilenberger-Sol-inter-Gap-Delta1-Qnot0}), the dependence of the gap on $Q_{1,x}$ and
$Q_{1,y}$ is qualitatively similar, although not exactly the same. It can be also seen that the gap does
not change when the signs of $Q_{1,x}$ and $Q_{1,y}$ are flipped.
However, this is not the case when the sign of $Q_{1,z}$ is changed.
As one can see from Eq.~(\ref{Eilenberger-Sol-inter-Gap-Delta1-Qnot0}),
the corresponding nontrivial dependence is rooted in the terms proportional to $\chi Q_{1,z} B_5$.
This finding is also supported by the numerical results in the right panel of
Fig.~\ref{fig:Eilenberger-Sol-inter-Gap-Delta1-B5}, where the value of
$|\Delta_1|$ is presented as a function of $B_5$ for several fixed values of $Q_{1,z}$.
At sufficiently small values of $B_5$, the terms quadratic in $Q_{1,z}$ have the tendency to suppress $|\Delta_1|$.
On the other hand, for sufficiently large pseudomagnetic fields, the gap can be either an increasing (if $\chi Q_{1,z}<0$)
or decreasing (if $\chi Q_{1,z}>0$) function of $Q_{1,z}$. Also, in the case when $\chi Q_{1,z}<0$,
we find that the value of the critical field strength $B_5^{\rm crit}$ can become larger than that at $Q_{1,z}=0$ (see also Eq.~(\ref{Eilenberger-Sol-inter-Gap-Delta1-B5-crit})). Briefly summarizing the obtained results, we found that the pseudomagnetic field tends to inhibit the spin-triplet gap. This effect can be partially mitigated when the phase vector is antiparallel to the pseudomagnetic field. All other configurations of the phase vector decrease the gap and lead to a smaller critical field strength.

\begin{figure}[t]
\begin{center}
\includegraphics[width=0.47\textwidth]{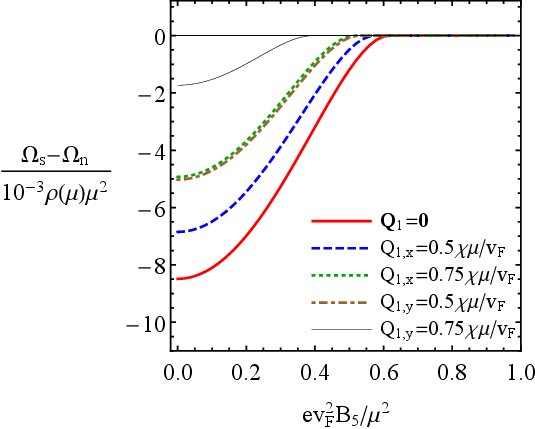}\hfill
\includegraphics[width=0.47\textwidth]{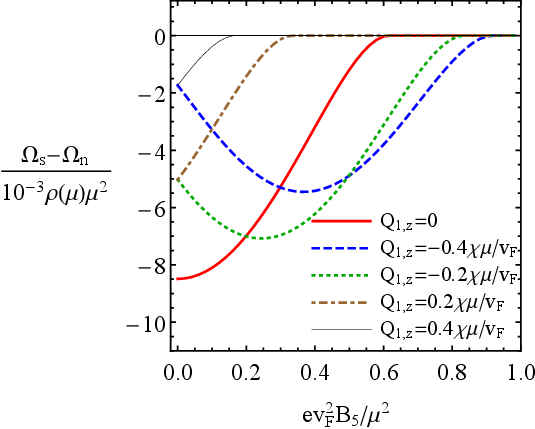}
\caption{The dependence of the energy densities difference between the superconducting and normal states $\Omega_{\rm s} -\Omega_{\rm n}$ given in
Eq.~(\ref{Eilenberger-Sol-inter-Gap-Delta1-Omega-def}) on the pseudomagnetic field strength $B_5$
for several values of the phase vector components $Q_{1,x}$ and $Q_{1,y}$ (left panel), and $Q_{1,z}$ (right panel).
Generically, only the phase vector component $Q_{1,z}$ can be energetically favorable at nonzero pseudomagnetic field if $\chi Q_{1,z}<0$.
To plot the results we used $T=0.45\mu$ and $T_0=0.5\mu$.}
\label{fig:Eilenberger-Sol-inter-Gap-Delta1-Omega-B5}
\end{center}
\end{figure}

In order to determine the ground state of the system, it is necessary to compare the energy
densities of the superconducting and normal states. The difference between the corresponding
energies can be expressed as follows \cite{Landau:t9}:
\begin{equation}
\label{Eilenberger-Sol-inter-Gap-Delta1-Omega-def}
\Omega_{\rm s} -\Omega_{\rm n} = -\int_{0}^{|U|} d|U| \frac{|\Delta_{1}|^2}{|U|^2}
= -\frac{2}{3}\rho(\mu) \int_{0}^{T_0} dT_0 \frac{|\Delta_{1}|^2}{T_0},
\end{equation}
where we used Eq.~(\ref{Eilenberger-Sol-inter-Gap-Delta1-Tc-def}) to obtain the last expression.
The numerical results for the energy densities difference (\ref{Eilenberger-Sol-inter-Gap-Delta1-Omega-def})
between the superconducting and normal phases are presented in
Fig.~\ref{fig:Eilenberger-Sol-inter-Gap-Delta1-Omega-B5} for several choices of phase
vectors $\mathbf{Q}_1$. In agreement with the results for the gap in
Fig.~\ref{fig:Eilenberger-Sol-inter-Gap-Delta1-B5}, the presence of $Q_{1,x}$ and $Q_{1,y}$ is
always unfavorable. On the other hand, the situation with $Q_{1,z}\neq0$ is different.
While at small $B_5$ the state with a nonzero $Q_{1,z}$ is also unfavorable, the superconducting
state with the phase satisfying $\chi Q_{1,z}<0$ has a lower energy at some nonzero $B_5$.

It should be noted that the energy density difference between the superconducting and normal
states can be estimated qualitatively by calculating the square of $|\Delta_1|$ in Eq.~(\ref{Eilenberger-Sol-inter-Gap-Delta1-Qnot0}), see also Fig.~\ref{fig:Eilenberger-Sol-inter-Gap-Delta1-B5}.
By making use of such an approach, we can determine the value of $\mathbf{Q}_1$ that
corresponds to the maximum gap $|\Delta_1|$ and, therefore, the energetically most favorable
superconducting state. The condition of a local extremum for the gap function is given by
the system of equations $\partial |\Delta_1|^2/\partial Q_{1,i}=0$, where $i=x,y,z$. Its
solution reads
\begin{eqnarray}
\label{Eilenberger-Sol-inter-Gap-Delta1-Q1-extr-xy}
Q_{1,x}^{\rm extr} &=& Q_{1,y}^{\rm extr}=0,\\
\label{Eilenberger-Sol-inter-Gap-Delta1-Q1-extr-z}
Q_{1,z}^{\rm extr} &\approx& -35\chi v_F eB_5 \mu\frac{49 \left(\zeta(3)\right)^3 T_0 -186\zeta(5)(T_0-T)}{70\zeta(3)\pi^2T^2(T_0-T) +4\mu^2 \left[343 \left(\zeta(3)\right)^3 T_0 -1395 \zeta(5) (T_0-T)\right]} +O(B_5^2).
\end{eqnarray}
[Note that there is another solution, but it has parametrically large (determined by $\mu$) value lying outside the validity
range of the model and, therefore, should be omitted.] By recalling the similarity between the gap
functions $\Delta_1$ and $\Delta_2$, it is clear that
the result for $\mathbf{Q}_2^{\rm extr}$ should be exactly the same. As for the phase vector in
the gap function $\Delta_3$, it should be given by a similar expression but with the vanishing
$B_5$. By noting that the result in Eq.~(\ref{Eilenberger-Sol-inter-Gap-Delta1-Q1-extr-z}) is
trivial in the limit $B_5\to0$, we conclude that $\mathbf{Q}_3^{\rm extr}=\mathbf{0}$.

The numerical results for the absolute value of gap $|\Delta_1|$, the difference of the energy
densities $\Omega_{\rm s} -\Omega_{\rm n}$, and the phase diagram in $T$--$B_5$ plane are
shown in three panels of Fig.~\ref{fig:Eilenberger-Sol-inter-Gap-Delta1-Omega-B5-Extr} for two different choices of the phase vector $\mathbf{Q}_1=\mathbf{0}$ and $\mathbf{Q}_1=\mathbf{Q}_{1}^{\rm extr}$, where the latter is given by Eqs.~(\ref{Eilenberger-Sol-inter-Gap-Delta1-Q1-extr-xy}) and (\ref{Eilenberger-Sol-inter-Gap-Delta1-Q1-extr-z}).
As expected, the solution with the phase $\mathbf{Q}_1=\mathbf{Q}_{1}^{\rm extr}$ has a
lower energy and a larger critical value of the pseudomagnetic field strength $B_5^{\rm crit}$. Physically this means that the corresponding superconducting state in the presence of pseudomagnetic fields should be spatially modulated.
This result also agrees qualitatively with the findings of Ref.~\cite{Matsushita:2018bty}, where a spatial modulation of a gap was advocated for a $p$-wave superconducting state.

Before concluding this subsection, let us briefly summarize our main results here. We found
that the pseudomagnetic field $\mathbf{B}_5$ inhibits the spin-triplet gaps when the
spins of Cooper pairs are normal to the field.
This is not the case when the spins are parallel to
$\mathbf{B}_5$, where the pseudomagnetic field neither affects the value of the gap nor leads to its spatial modulation.
In agreement with an earlier study in Ref.~\cite{Matsushita:2018bty}, the spatial modulation
of the gap, which is determined by the phase vector $\mathbf{Q}_{1}$, leads to an energetically
more favorable state for $\Delta_1\neq0$ or $\Delta_2\neq0$.
Such a state, however, still has a higher energy than that with $\Delta_3\neq0$.

Finally, let us briefly discuss the case where the pseudomagnetic fields are absent and the spatial gradients can be ignored. 
Here, the fully gapped spin-singlet superconducting state with the intra-node pairing
might have a higher critical temperature and could be more energetically favorable than the spin-triplet
states with the inter-node pairing. This suggests that there might exist a critical value of the
pseudomagnetic field strength separating the spin-singlet state with the intra-node pairing from the
spin-triplet one with the inter-node pairing. By taking into account, however, that the former
should be spatially nonuniform in the presence of a nonzero $\mathbf{B}_5$, the study
of the competition between these two phases is a nontrivial task. While it is beyond the
scope of this study, it should be addressed in future investigations.

\subsection{Currents}
\label{sec:Eilenberger-Sol-inter-Currents}

Since the superconducting gaps generically have nontrivial phase factors, one might expect
that nonzero electric and/or chiral currents could be induced in the corresponding states of
strained Weyl semimetals. (Note also that strain can be used to control supercurrents in some two dimensional systems \cite{Alidoust:2018}.)
The formal expression for the current in each chiral sector is
given by
\begin{equation}
\label{Eilenberger-Sol-inter-Currents-current-chi-def}
\mathbf{j}_{\chi} \approx 2e\rho(\mu) \pi  iT\sum_{m^{\prime}=-\infty}^{\infty} \int \frac{d\Omega_{\mathbf{k}^{\prime}_{\parallel} }}{4\pi} v_F \hat{\mathbf{k}}^{\prime} \left[g_0(\mathbf{k}^{\prime}_{\parallel} ,\mathbf{R})+g_1(\mathbf{k}^{\prime}_{\parallel} ,\mathbf{R})+g_2(\mathbf{k}^{\prime}_{\parallel} ,\mathbf{R})\right],
\end{equation}
which is valid up to the second order in external fields and spatial derivatives.
For concreteness, let us consider the contribution due to the single gap component $\Delta_1$.
By taking into account the similarity of the gap equations for the other components, of course, the
analysis could be straightforwardly performed in the case of nonzero $\Delta_2$ and
$\Delta_3$.
(In this connection, we should remind that, in addition to the replacements of the phase vectors, $B_5$ should be set to zero in the case of $\Delta_{3}\neq0$.)
From the definition in Eq.~(\ref{Eilenberger-Sol-inter-Currents-current-chi-def}),
we derive the following expression for the current:
\begin{equation}
\label{Eilenberger-Sol-inter-Currents-Delta1-current-chi}
\mathbf{j}_{\chi}
= -\frac{7\zeta(3)v_F^2 e \rho(\mu) |\Delta_1|^2}{12\pi^2T^2} \left[ \frac{4}{5} \left(\mathbf{Q}_1 -\frac{1}{2}\hat{\mathbf{x}}Q_{1,x}\right) +\chi \frac{v_F eB_5}{\mu}\hat{\mathbf{z}}\right],
\end{equation}
where we used the formula in Eq.~(\ref{Eilenberger-Sol-inter-Gap-Delta1-zeta-def}) and took into
account that the zeroth order contribution vanishes after the Matsubara summation. By making use
of this result and substituting the phase
vector $\mathbf{Q}_1 =\mathbf{Q}_{1}^{\rm extr}$, where $\mathbf{Q}_{1}^{\rm extr}$ is given by
Eqs.~(\ref{Eilenberger-Sol-inter-Gap-Delta1-Q1-extr-xy}) and (\ref{Eilenberger-Sol-inter-Gap-Delta1-Q1-extr-z}),
we obtain the following results for the electric and chiral currents:
\begin{eqnarray}
\label{Eilenberger-Sol-inter-Currents-Delta1-current-e}
\mathbf{j}_e &=& \sum_{\chi=\pm}\mathbf{j}_{\chi} = \mathbf{0},\\
\label{Eilenberger-Sol-inter-Currents-Delta1-current-5}
\mathbf{j}_5 &=& \sum_{\chi=\pm}\chi\mathbf{j}_{\chi} = -\frac{7\zeta(3)v_F^2 e \rho(\mu) |\Delta_1|^2}{6\pi^2T^2} \left( \frac{4Q_{1,z}^{\rm extr}}{5\chi} +\frac{v_F eB_5}{\mu} \right) \hat{\mathbf{z}}.
\end{eqnarray}
Here we took into account that $Q_{1,z}^{\rm extr}$ given in Eq.~(\ref{Eilenberger-Sol-inter-Gap-Delta1-Q1-extr-z}) is proportional to $\chi$.
As is clear, the currents in the case with $\Delta_2\neq0$ will be exactly the same (up to the
replacement of $|\Delta_1|$ with $|\Delta_2|$ and $\mathbf{Q}_{1}^{\rm extr}$ with $\mathbf{Q}_{2}^{\rm extr}$).
On the other hand, in the case of a nonzero $\Delta_3$,
i.e., when the spins of Cooper pairs are parallel to the pseudomagnetic field, both electric and chiral
currents are absent.

\begin{figure}[t]
\begin{center}
\includegraphics[width=0.45\textwidth]{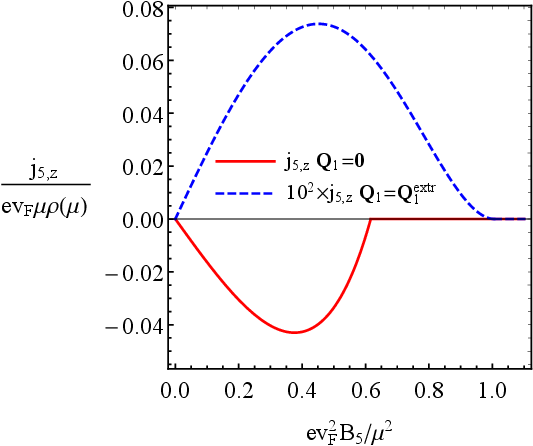}
\caption{The dependence of the chiral current component $j_{5,z}$ given by Eq.~(\ref{Eilenberger-Sol-inter-Currents-Delta1-current-5}) on the pseudomagnetic field strength $B_5$ for the extremal value of the phase vector $\mathbf{Q}_1=\mathbf{Q}_{1}^{\rm extr}$ provided in Eqs.~(\ref{Eilenberger-Sol-inter-Gap-Delta1-Q1-extr-xy}) and (\ref{Eilenberger-Sol-inter-Gap-Delta1-Q1-extr-z}).
To plot the results we used $T=0.45\mu$ and $T_0=0.5\mu$.}
\label{fig:Eilenberger-Eqs-sol-inter-study-Delta1-current-5}
\end{center}
\end{figure}

The dependence of the $z$ component of the chiral current
(\ref{Eilenberger-Sol-inter-Currents-Delta1-current-5}) on the pseudomagnetic field strength $B_5$ is
presented in Fig.~\ref{fig:Eilenberger-Eqs-sol-inter-study-Delta1-current-5}. The current is a nonmonotonous function that is determined by an interplay of gap $|\Delta_1|$, which decreases with $B_5$, and a linear in $B_5$ term. By comparing Figs.~\ref{fig:Eilenberger-Sol-inter-Gap-Delta1-B5} and \ref{fig:Eilenberger-Eqs-sol-inter-study-Delta1-current-5}, it is evident that the current vanishes simultaneously with the superconducting gap.
As we see, the chiral current is considerably smaller in the state with $\mathbf{Q}_1=\mathbf{Q}_{1}^{\rm extr}$
than in the spatially uniform state. Technically, this is due to the
fact that the two contributions in the parentheses in
Eq.~(\ref{Eilenberger-Sol-inter-Currents-Delta1-current-5}) are comparable and largely
compensate each other. In view of this, it might be tempting to suggest then that the approximate value
of the phase vector was not determined precisely enough and the correct result
should give a vanishing $\mathbf{j}_5$. We checked, however, that the
value of $\mathbf{Q}_1$ that enforces the condition $\mathbf{j}_5=\mathbf{0}$
leads to a higher energy state.

\section{Summary}
\label{sec:Summary}

In this study, by using the quasiclassical Eilenberger approach, we studied the effects of the strain-induced
pseudomagnetic field $\mathbf{B}_5$ on the inter-node spin-triplet superconductivity in Weyl semimetals with a broken TR
symmetry. Unlike the usual magnetic field, the pseudomagnetic one couples to fermions from the Weyl nodes of opposite chirality with different sign. This immediately leads to an unusual dynamics of Cooper pairs, where the Meissner effect is absent.
In agreement with previous studies, we found that only the spin-triplet channel is relevant for
the inter-node pairing of quasiparticles in the vicinity of the Fermi surface (i.e., the pairing of quasiparticles from the Weyl nodes of opposite chirality).
As expected, when the pseudomagnetic field is absent, there is no preferred direction for the spins
of Cooper pairs and the spatial modulation of the superconducting gap is not favorable.
The dynamics changes qualitatively in the presence of the pseudomagnetic field.
The superconducting state with the spins of Cooper pairs parallel to $\mathbf{B}_5$ is
not affected by the field and has the lowest energy among the spin-triplet states.
On the other hand, the states with the spins normal to $\mathbf{B}_5$ are modified and inhibited by the field.
In the latter case, the gap function acquires a nontrivial phase proportional to the field strength, which lowers the energy of this state.
This result qualitatively agrees with that in Ref.~\cite{Matsushita:2018bty}. Thus, the pseudomagnetic field effectively selects one of the spin-triplet superconducting states by inhibiting the other two.

We also investigated the currents in the superconducting ground state. While the electric currents
always vanish, the spin-triplet state could give rise to a nonzero chiral current $\mathbf{j}_5$, which is
determined by the pseudomagnetic field and the phase of the gap. From a physical point of view, this current corresponds to a spin polarization that could be, in principle, detected experimentally.
Interestingly, for the Cooper pairing with spins normal to the direction of the pseudomagnetic field, $\mathbf{j}_5$
is nonzero, albeit small for the energetically most favorable phase vector. The chiral current is completely absent
when the spins of Cooper pairs are parallel to $\mathbf{B}_5$. Since the latter has the lowest
energy, we believe that both electric and chiral supercurrents are absent in the spin-triplet ground
state of Weyl semimetals with the inter-node superconducting pairing.

At the end, let us briefly discuss the limitations of this study and outlook for future
investigations. In our analysis, we used a rather simple model of Weyl semimetals with a
broken TR symmetry. While the key qualitative results are likely to remain valid also
for realistic materials with multiple pairs of Weyl nodes, it would be interesting to verify this
by a direct analysis. Another limitation of this study is connected with the use of the
quasiclassical Eilenberger equation whose validity is restricted to weak pseudomagnetic
fields. It would be important, therefore, to rigorously investigate the role of strong pseudomagnetic fields on the superconductivity when the pseudo-Landau levels are formed. The use of the Ginzburg--Landau theory implies
that our study becomes unreliable far away from the superconducting phase transition where
the gap is not very small. Such a limitation should be overcome in the future investigations by
employing more sophisticated approaches. They include, for example, the Eliashberg method \cite{Eliashberg:1960} for the Green's functions in the background pseudomagnetic field or the functional renormalization group approach, where
the smallness of the gap is not required and the gradient expansion can be avoided. In this connection, it should
be noted that the assumption of the smallness of the gap is not required even in the semiclassical Eilenberger
approach used in this study. Without it, however, the analysis would become significantly more complicated.
Last but not least, in this paper, we studied primarily the inter-node pairing of quasiparticles from different Weyl nodes.
On the other hand, the case of intra-node pairing, which was briefly discussed at the beginning of Sec.~\ref{sec:Eilenberger-Eqs-sol} as well as at the end of Sec.~\ref{sec:Eilenberger-Sol-inter-Gap}, could result in rather complicated, spatially
nonuniform solutions and deserve an in-depth investigation. The corresponding task is highly nontrivial and is beyond the scope of this paper.

\begin{acknowledgments}
The work of E.V.G. was partially supported by the Program of Fundamental Research of the
Physics and Astronomy Division of the National Academy of Sciences of Ukraine.
The work of V.A.M. and P.O.S. was supported by the Natural Sciences and Engineering Research Council of Canada.
The work of I.A.S. was supported by the U.S. National Science Foundation under Grant PHY-1713950.
\end{acknowledgments}

\appendix

\section{Expansion of the Gor'kov equations}
\label{sec:App-expansion}

In this appendix, we expand the Gor'kov equations (\ref{Model-Gor'kov-Eq-New-FT-1}) and (\ref{Model-Gor'kov-Eq-New-FT-2})
up to the second order in spatial gradients and the uniform chiral magnetic field $\mathbf{B}_{\chi}$.

The left-hand side of the first Gor'kov equation (\ref{Model-Gor'kov-Eq-New-FT-1}) gives
\begin{eqnarray}
\label{Model-Gor'kov-Eq-FT-1-all}
&&\left[-i\omega_{m^{\prime}} \tau_z +\tilde{H} \left(\mathbf{k} +\tau_z e \mathbf{A}_{\chi}(\mathbf{R}) \right) +\tilde{\Delta}(\mathbf{k}, \mathbf{R})\right] \circ \tilde{G}(\mathbf{k}, \mathbf{R}; i\omega_{m^{\prime}}) \approx -i\omega_{m^{\prime}} \tau_z\tilde{G} +\tilde{H} \tilde{G} + \tau_z e A_{\chi}^{j}  (\partial_{k_{j}} \tilde{H}) \tilde{G}\nonumber\\
&&+\frac{e^2}{2} A_{\chi}^{j}A_{\chi}^{l} (\partial_{k_{j}} \partial_{k_{l}} \tilde{H}) \tilde{G} +\tau_z \frac{ie}{2} (\partial_{k_{l}} \tilde{H})
\epsilon^{jlm}B_{\chi}^{m}
(\partial_{k_{j}} \tilde{G})
+\frac{ie^2}{2} A_{\chi}^{l} (\partial_{k_{l}} \partial_{k_{m}}\tilde{H})
\epsilon^{jmn}B_{\chi}^{n}
(\partial_{k_{j}} \tilde{G}) \nonumber\\
&&-\frac{i}{2} (\partial_{k_{j}}\tilde{H}) (\partial_{R_{j}}\tilde{G})
-\tau_z\frac{ie}{2}A_{\chi}^{l} (\partial_{k_{j}} \partial_{k_{l}} \tilde{H}) (\partial_{R_{j}}\tilde{G})
-\frac{1}{8} (\partial_{k_{j}} \partial_{k_{l}}\tilde{H}) (\partial_{R_{j}} \partial_{R_{l}} \tilde{G})
+\tau_z\frac{e}{4}
\epsilon^{lmn}B_{\chi}^n
(\partial_{k_m} \partial_{k_{j}}\tilde{H}) (\partial_{k_{l}} \partial_{R_{j}} \tilde{G}) \nonumber\\
&&-\frac{e^2}{8}
\epsilon^{lmp}B_{\chi}^{p} \epsilon^{jns}B_{\chi}
(\partial_{k_{m}} \partial_{k_{n}} \tilde{H}) (\partial_{k_{j}} \partial_{k_{l}} \tilde{G})
+\tilde{\Delta} \tilde{G}
+\frac{i}{2} \left[(\partial_{R_{j}} \tilde{\Delta})(\partial_{k_{j}} \tilde{G}) -(\partial_{k_{j}} \tilde{\Delta})(\partial_{R_{j}}\tilde{G})\right]
\nonumber\\
&&
-\frac{1}{8} (\partial_{R_{j}}\partial_{R_{l}} \tilde{\Delta}) (\partial_{k_{j}}\partial_{k_{l}} \tilde{G}) +\frac{1}{4}(\partial_{R_{j}}\partial_{k_{l}} \tilde{\Delta}) (\partial_{k_{j}}\partial_{R_{l}} \tilde{G})
-\frac{1}{8} (\partial_{k_{j}}\partial_{k_{l}} \tilde{\Delta}) (\partial_{R_{j}}\partial_{R_{l}} \tilde{G})
+O\left(\partial_{R_{j}}^3, (A_{\chi}^{j})^3, (\partial_{j} A_{\chi}^{l})^3\right).
\end{eqnarray}
Here $\tilde{H}$, $\tilde{\Delta}$, and $\tilde{G}$ are the transformed Bogolyubov--de Gennes (BdG) Hamiltonian, the gap matrix, and the Green's function defined in Eqs.~(\ref{Model-Gor'kov-new-vars-H}), (\ref{Model-Gor'kov-new-vars-Delta}), and (\ref{Model-Gor'kov-new-vars-G}) in the main text, respectively. In addition, we used the definition of the circle product in Eq.~(\ref{Model-Gor'kov-circle-def}), $\tau_z$ is the Pauli matrix acting in the Nambu--Gor'kov space, $\omega_{m^{\prime}}$ is the Matsubara frequency, $e$ is the electron charge, $\mathbf{k}$ is the momentum, $\mathbf{R}$ is the centre-of-mass coordinate, and $\epsilon^{jlm}$ is the antisymmetric Levi-Civita tensor. The expansion for the second Gor'kov equation (\ref{Model-Gor'kov-Eq-New-FT-2}) can be performed in a similar way.

In order to obtain a homogeneous equation for the Green's function $\tilde{G}(\mathbf{k}, \mathbf{R}; i\omega_{m^{\prime}})$, we subtract the first and second expanded Gor'kov equations. The final result reads
\begin{eqnarray}
\label{Model-Gor'kov-Eilenberger-Eq-diff}
&&-i\omega_{m^{\prime}} \left[\tau_z, \tilde{G}\right] + \left[\tilde{H}, \tilde{G}\right]
+ eA_{\chi}^{j}\left[\tau_z (\partial_{k_{j}} \tilde{H}), \tilde{G}\right]
+ \frac{i e}{2} \epsilon^{jlm}B_{\chi}^{m}
\left\{\tau_z (\partial_{k_{l}}\tilde{H}), (\partial_{k_{j}}\tilde{G})\right\}
-\frac{i}{2} \left\{(\partial_{k_{j}}\tilde{H}), (\partial_{R_{j}}\tilde{G})\right\} \nonumber\\
&&+\frac{e^2}{2}A^{j}_{\chi} A^{l}_{\chi}\left[(\partial_{k_{j}} \partial_{k_{l}}\tilde{H}), \tilde{G}\right]
+\frac{ie^2}{2} A^{l}_{\chi} \epsilon^{j m n}B_{\chi}^{n}
\left\{(\partial_{k_{m}}\partial_{k_{l}} \tilde{H}), (\partial_{k_{j}} \tilde{G}) \right\}
-\frac{ie}{2} A^{l}_{\chi} \left\{\tau_z (\partial_{k_{j}}\partial_{k_{l}} \tilde{H}), (\partial_{R_{j}} \tilde{G}) \right\} \nonumber\\
&&-\frac{e^2}{8} \epsilon^{l m p} B_{\chi}^{p} \epsilon^{j n s} B_{\chi}^{s} \left[(\partial_{k_{m}}\partial_{k_{n}} \tilde{H}), (\partial_{k_{j}}\partial_{k_{l}} \tilde{G})\right]
+\frac{e}{4} \epsilon^{l m n} B_{\chi}^{n} \left[\tau_z(\partial_{k_{m}}\partial_{k_{j}}\tilde{H}), (\partial_{k_{l}}\partial_{R_{j}} \tilde{G})\right]
\nonumber\\
&&-\frac{1}{8} \left[(\partial_{k_{j}}\partial_{k_{l}} \tilde{H}), (\partial_{R_{j}}\partial_{R_{l}} \tilde{G}) \right]
+\left[\tilde{\Delta}, \tilde{G}\right]
+\frac{i}{2} \left\{(\partial_{R_{j}} \tilde{\Delta}), (\partial_{k_{j}} \tilde{G}) \right\}
-\frac{i}{2} \left\{(\partial_{k_{j}}\tilde{\Delta}), (\partial_{R_{j}}\tilde{G}) \right\}
\nonumber\\
&&
-\frac{1}{8} \left[(\partial_{R_{j}}\partial_{R_{l}} \tilde{\Delta}), (\partial_{k_{j}}\partial_{k_{l}}\tilde{G}) \right]
+\frac{1}{4} \left[(\partial_{k_{j}}\partial_{R_{l}}\tilde{\Delta}), (\partial_{R_{j}}\partial_{k_{l}}\tilde{G})\right]
-\frac{1}{8} \left[(\partial_{k_{j}}\partial_{k_{l}} \tilde{\Delta}), (\partial_{R_{j}}\partial_{R_{l}}\tilde{G})\right]
\nonumber\\
&& +O\left(\partial_{R_{j}}^3, (A_{\chi}^{j})^3, (\partial_{j} A_{\chi}^{l})^3\right)=0.
\end{eqnarray}
Here, for the sake of brevity, we omitted the arguments
$\mathbf{k}$ and $\mathbf{R}$ in functions $\tilde{G}$, $\tilde{H}$, and $\tilde{\Delta}$. In addition,
the square and curly brackets denote the commutators and anticommutators, respectively.

\section{Integrated Green's functions for inter-node pairing}
\label{sec:Eilenberger-sol-inter}

In this appendix, we present the details of iterative solution to the Eilenberger equation (\ref{Model-Gor'kov-Eilenberger-Eq-final}) amended with the normalization condition (\ref{Model-Gor'kov-Eilenberger-g-norm-Eq}). The integrated Green's function $\tilde{g}$, defined in Eq.~(\ref{Eilenberger-Eq-Green-int}), is expanded  up to the second order in powers of the background fields and spatial derivatives, see Eq.~(\ref{Eilenberger-sol-inter-g-expand}) in the main text.
By making use of the reduced BdG Hamiltonian (\ref{Model-BdG-reduction-HBdG-inter-2reduced}) for the inter-node pairing, as well as
the definitions in Eqs.~(\ref{Model-Gor'kov-new-vars-H}), (\ref{Model-Gor'kov-Eilenberger-V-def}), and
(\ref{Model-Gor'kov-Eilenberger-M-def}), we derive the following explicit expressions for $\tilde{\mathbf{V}}$
and $\tilde{M}_{jl}$:
\begin{eqnarray}
\label{Eilenberger-sol-inter-V}
\tilde{\mathbf{V}} &=& v_F \hat{\mathbf{k}},\\
\label{Eilenberger-sol-inter-M}
\tilde{M}_{jl} &=& \frac{v_F^2}{\mu}\left(\delta_{jl} - \hat{\mathbf{k}}_{j}\hat{\mathbf{k}}_{l}\right),
\end{eqnarray}
where $v_F$ is the Fermi velocity, $\mu$ is the electric chemical potential, and, for the sake of simplicity, we omitted the chirality index $\chi$ at $\mathbf{k}$.

As discussed at the beginning of Sec.~\ref{sec:Eilenberger-Eqs-sol}, it is convenient to assume that the ordinary magnetic field is absent, i.e., $\mathbf{A}_{\chi}=\chi\mathbf{A}_5$ and $\mathbf{B}_{\chi}=\chi\mathbf{B}_5$.
By recalling that the axial vector potential is proportional to the chirality of Weyl nodes, we should also replace $\mathbf{A}_{5} \to \tau_z\mathbf{A}_{5}$ and $\mathbf{B}_{5} \to \tau_z\mathbf{B}_{5}$ in Eq.~(\ref{Model-Gor'kov-Eilenberger-Eq-final}) for the case of inter-node pairing.

\subsection{Zeroth order}
\label{sec:Eilenberger-sol-inter-Zero}

Let us start from the solution in the zeroth order approximation, in which background fields and spatial
gradients are neglected. The corresponding Eilenberger equation reads
\begin{equation}
\label{Eilenberger-Sol-inter-Zero-eq}
-i\omega_{m^{\prime}} \left[\tau_z, \tilde{g}_0\right] + \left[\tilde{\Delta}, \tilde{g}_0\right] =0.
\end{equation}
By taking into account the matrix structure of $\tilde{g}_0$, see Eq.~(\ref{Eilenberger-Eq-Green-int}) in the main text, it is straightforward to show that
there are two nontrivial equations for the components of $\tilde{g}_0$. Their solution reads
\begin{eqnarray}
\label{Eilenberger-Sol-inter-Zero-rel-f0dag}
f_0^{\dag} &=& f_0 \frac{\Delta_{\chi}^{*}}{\Delta_{\chi}},\\
\label{Eilenberger-Sol-inter-Zero-rel-g0bar}
\bar{g}_0 &=& g_0- \frac{2i\omega_{m^{\prime}} f_0}{\Delta_{\chi}},
\end{eqnarray}
where $\Delta_{\chi}$ is given by Eq.~(\ref{Model-BdG-reduction-Delta-chi-inter-def}). In order to determine
$g_0$ and $f_0$, one needs to employ the normalization condition (\ref{Model-Gor'kov-Eilenberger-g-norm-Eq}),
which in the zeroth order is equivalent to
\begin{eqnarray}
\label{Eilenberger-Sol-inter-Zero-norm-1}
&&\bar{g}_0 = -g_0,\\
\label{Eilenberger-Sol-inter-Zero-norm-2}
&&g_0^2-f_0f_0^{\dag} = 1.
\end{eqnarray}
By taking these conditions into account, we obtain
\begin{eqnarray}
\label{Eilenberger-Sol-inter-Zero-sol-f0}
f_0 &=& - (f_0^{\dag})^{*}=-\frac{i\Delta_{\chi}}{\sqrt{\omega_{m^{\prime}}^2 +|\Delta_{\chi}|^2}},\\
\label{Eilenberger-Sol-inter-Zero-sol-g0}
g_0 &=& -\bar{g}_0 =\frac{\omega_{m^{\prime}}}{\sqrt{\omega_{m^{\prime}}^2 +|\Delta_{\chi}|^2}}.
\end{eqnarray}
In passing, let us mention that the same relations would be also valid in the case of intra-node pairing with the corresponding value of the gap.

\subsection{First order}
\label{sec:Eilenberger-sol-inter-First}

In the first order in background fields and spatial gradients, the Eilenberger equation
(\ref{Model-Gor'kov-Eilenberger-Eq-final}) for the inter-node pairing reads
\begin{eqnarray}
\label{Eilenberger-Sol-inter-First-eq}
&&-i\omega_{m^{\prime}} \left[\tau_z, \tilde{g}_1\right]+ i e v_F \left(\left[\hat{\mathbf{k}}\times\mathbf{B}_{\chi}\right] \cdot \bm{\nabla}_{\mathbf{k}_{\parallel}}\right)\tilde{g}_0
-iv_F \left(\hat{\mathbf{k}}\cdot \bm{\nabla}_{\mathbf{R}}\right) \tilde{g}_0
\nonumber\\
&&+\left[\tilde{\Delta}, \tilde{g}_1\right]
+\frac{i}{2} \left\{(\bm{\nabla}_{\mathbf{R}}\tilde{\Delta}), (\bm{\nabla}_{\mathbf{k}_{\parallel}}\tilde{g}_0)\right\}
-\frac{i}{2} \left\{(\bm{\nabla}_{\mathbf{k}_{\parallel}} \tilde{\Delta}), (\bm{\nabla}_{\mathbf{R}}\tilde{g}_0)\right\}
=0.
\end{eqnarray}
As in the leading order approximation, the matrix equation (\ref{Eilenberger-Sol-inter-First-eq}) contains only two nontrivial equations that produce the following relations:
\begin{equation}
\label{Eilenberger-Sol-inter-First-rel-fdag}
f_1^{\dag} = f_1 \frac{\Delta^{*}_{\chi}}{\Delta_{\chi}} -\frac{iv_F}{\Delta_{\chi}}\left\{e\left(\left[\hat{\mathbf{k}}\times\mathbf{B}_{\chi}\right]\cdot \bm{\nabla}_{\mathbf{k}_{\parallel}}\right) -(\hat{\mathbf{k}}\cdot\bm{\nabla}_{\mathbf{R}})\right\} g_0
\end{equation}
and
\begin{equation}
\label{Eilenberger-Sol-inter-First-rel-gbar}
\bar{g}_1 = g_1
-\frac{2i\omega_{m^{\prime}} (f_1+f_1^{\dag})}{\Delta_{\chi}+\Delta_{\chi}^{*}} +\frac{iv_F}{\Delta_{\chi}+\Delta_{\chi}^{*}}\left\{e\left(\left[\hat{\mathbf{k}}\times\mathbf{B}_{\chi}\right]\cdot \bm{\nabla}_{\mathbf{k}_{\parallel}}\right) -(\hat{\mathbf{k}}\cdot\bm{\nabla}_{\mathbf{R}})\right\}
(f_0-f_0^{\dag}).
\end{equation}
Here the normalization conditions for the zeroth order solutions in Eqs.~(\ref{Eilenberger-Sol-inter-Zero-norm-1}) and (\ref{Eilenberger-Sol-inter-Zero-norm-2}) were used.
By retaining the first order terms in the normalization condition (\ref{Model-Gor'kov-Eilenberger-g-norm-Eq}), we obtain
\begin{eqnarray}
\label{Eilenberger-Sol-inter-First-norm-1}
&&\bar{g}_1 = -g_1,\\
\label{Eilenberger-Sol-inter-First-norm-2}
&&2g_0g_1 -f_1f_0^{\dag} -f_0f_{1}^{\dag} =0.
\end{eqnarray}
Taking these expressions into account, we derive the following anomalous and normal components of the integrated
Green's function:
\begin{eqnarray}
\label{Eilenberger-Sol-inter-First-sol-f1}
f_{1} &=& -i\frac{v_F f_0}{2\Delta_{\chi}} \left(f_0 -\frac{2i\omega_{m^{\prime}}g_0}{\Delta_{\chi}+\Delta_{\chi}^{*}}\right) \left\{e\left(\left[\hat{\mathbf{k}}\times\mathbf{B}_{\chi}\right]\cdot \bm{\nabla}_{\mathbf{k}_{\parallel}}\right) -(\hat{\mathbf{k}}\cdot\bm{\nabla}_{\mathbf{R}})\right\}g_0
\nonumber\\
&+&\frac{i v_F f_0g_0}{2\left(\Delta_{\chi}+\Delta_{\chi}^{*}\right)} \left\{e\left(\left[\hat{\mathbf{k}}\times\mathbf{B}_{\chi}\right]\cdot \bm{\nabla}_{\mathbf{k}_{\parallel}}\right) -(\hat{\mathbf{k}}\cdot\bm{\nabla}_{\mathbf{R}})\right\} \left(f_0\frac{\Delta_{\chi}-\Delta_{\chi}^{*}}{\Delta_{\chi}}\right)
\end{eqnarray}
and
\begin{eqnarray}
\label{Eilenberger-Sol-inter-First-sol-g1}
g_{1} &=& -\bar{g}_{1}= \frac{v_F \omega_{m^{\prime}} \left(\Delta_{\chi}-\Delta_{\chi}^{*}\right) f_0^2 }{2\Delta_{\chi}^2 \left(\Delta_{\chi}+\Delta_{\chi}^{*}\right)} \left\{e\left(\left[\hat{\mathbf{k}}\times\mathbf{B}_{\chi}\right]\cdot \bm{\nabla}_{\mathbf{k}_{\parallel}}\right) -(\hat{\mathbf{k}}\cdot\bm{\nabla}_{\mathbf{R}})\right\}g_0 \nonumber\\
&+&\frac{iv_F f_0^2 \Delta_{\chi}^{*}}{2\Delta_{\chi}\left(\Delta_{\chi}+\Delta_{\chi}^{*}\right)}\left\{e\left(\left[\hat{\mathbf{k}}\times\mathbf{B}_{\chi}\right]\cdot \bm{\nabla}_{\mathbf{k}_{\parallel}}\right) -(\hat{\mathbf{k}}\cdot\bm{\nabla}_{\mathbf{R}})\right\} \left(f_0\frac{\Delta_{\chi}-\Delta_{\chi}^{*}}{\Delta_{\chi}}\right),
\end{eqnarray}
respectively.
Here, $f^{\dag}_1$ can be obtained by using Eqs.~(\ref{Eilenberger-Sol-inter-First-rel-fdag}) and (\ref{Eilenberger-Sol-inter-First-sol-f1}).
It is worth noting that while the anomalous function $f_1$ in Eq.~(\ref{Eilenberger-Sol-inter-First-sol-f1})
is odd in $\omega_{m^{\prime}}$, its nonanomalous counterpart $g_1$ is an even function.
This property, as one can see in Sec.~\ref{sec:Eilenberger-Sol-inter-Gap}, is very important for solving the gap equation and for
calculating the superconducting currents.

\subsection{Second order}
\label{sec:Eilenberger-sol-inter-Second}

Next, we consider the second order in background fields and spatial derivatives. In such a case the equation for the integrated Green's function reads
\begin{eqnarray}
\label{Eilenberger-Sol-inter-Second-eq}
&&-i\omega_{m^{\prime}} \left[\tau_z, \tilde{g}_2\right]+ i e v_F\left(\left[\hat{\mathbf{k}}\times\mathbf{B}_{\chi}\right] \cdot \bm{\nabla}_{\mathbf{k}_{\parallel}}\right) \tilde{g}_1
-iv_F \left(\hat{\mathbf{k}}\cdot \bm{\nabla}_{\mathbf{R}}\right) \tilde{g}_1
+ie^2 A^{l}_{\chi} \epsilon^{j m n} B_{\chi}^{n} \tilde{M}_{ml} \partial_{k_{\parallel, j}} \tilde{g}_0
-ie A^{l}_{\chi} \tilde{M}_{j l} \partial_{R_{j}}\tilde{g}_0\nonumber\\
&&+\left[\tilde{\Delta}, \tilde{g}_2\right]
+\frac{i}{2} \left\{(\bm{\nabla}_{\mathbf{R}}\tilde{\Delta}), (\bm{\nabla}_{\mathbf{k}_{\parallel}} \tilde{g}_1) \right\}
-\frac{i}{2} \left\{(\bm{\nabla}_{\mathbf{k}_{\parallel}} \tilde{\Delta}), (\bm{\nabla}_{\mathbf{R}} \tilde{g}_1)\right\}
-\frac{1}{8}\left[(\partial_{R_{j}} \partial_{R_{l}} \tilde{\Delta}), (\partial_{k_{\parallel, j}} \partial_{k_{\parallel, l}} \tilde{g}_0)\right]
\nonumber\\
&&+\frac{1}{4}\left[(\partial_{R_{j}} \partial_{k_{\parallel, l}} \tilde{\Delta}), (\partial_{k_{\parallel, j}} \partial_{R_{l}} \tilde{g}_0)\right]
-\frac{1}{8}\left[(\partial_{k_{\parallel, j}} \partial_{k_{\parallel, l}} \tilde{\Delta}), (\partial_{R_{j}} \partial_{R_{l}} \tilde{g}_0)\right]
=0.
\end{eqnarray}
This matrix equation allows one to determine two out of the four components of $\tilde{g}_2$.
In particular, we derive
\begin{eqnarray}
\label{Eilenberger-Sol-inter-Second-eq-fdag2}
f_{2}^{\dag} &=& f_2\frac{\Delta^{*}_{\chi}}{\Delta_{\chi}}
-\frac{iv_F}{\Delta_{\chi}} \left\{e\left(\left[\hat{\mathbf{k}}\times\mathbf{B}_{\chi}\right]\cdot \bm{\nabla}_{\mathbf{k}_{\parallel}}\right) -(\hat{\mathbf{k}}\cdot\bm{\nabla}_{\mathbf{R}})\right\}g_1
-\frac{ie^2v_F^2}{\Delta_{\chi} \mu} \left(\left[\mathbf{A}_{\chi}\times \mathbf{B}_{\chi}\right]\cdot \bm{\nabla}_{\mathbf{k}_{\parallel}}\right)g_0 \nonumber\\
&+&\frac{ie^2v_F^2}{\Delta_{\chi} \mu} \left(\mathbf{A}_{\chi}\cdot \hat{\mathbf{k}}\right)\left(\left[\hat{\mathbf{k}}\times \mathbf{B}_{\chi}\right]\cdot \bm{\nabla}_{\mathbf{k}_{\parallel}}\right) g_0
+\frac{iev_F^2}{\Delta_{\chi} \mu} \left\{\left(\mathbf{A}_{\chi}\cdot\bm{\nabla}_{\mathbf{R}}\right) -\left(\mathbf{A}_{\chi}\cdot\hat{\mathbf{k}}\right) \left(\hat{\mathbf{k}}\cdot\bm{\nabla}_{\mathbf{R}}\right)\right\}g_0
-\frac{1}{2 \Delta_{\chi}}D_{\rm terms}^{(1)}
\end{eqnarray}
and
\begin{eqnarray}
\label{Eilenberger-Sol-inter-Second-eq-g2-gbar2}
\bar{g}_2&=& g_2+ \frac{1}{\Delta_{\chi}+\Delta_{\chi}^{*}} \Bigg\{-2i\omega_{m^{\prime}}(f_2^{\dag}+f_2)
+iv_F\left\{e\left(\left[\hat{\mathbf{k}}\times\mathbf{B}_{\chi}\right]\cdot \bm{\nabla}_{\mathbf{k}_{\parallel}}\right) -(\hat{\mathbf{k}}\cdot\bm{\nabla}_{\mathbf{R}})\right\}(f_1-f_1^{\dag})
\nonumber\\
&+& \frac{ie^2v_F^2}{\mu} \left\{\left(\left[\mathbf{A}_{\chi}\times \mathbf{B}_{\chi}\right]\cdot \bm{\nabla}_{\mathbf{k}_{\parallel}}\right) -\left(\mathbf{A}_{\chi}\cdot \hat{\mathbf{k}}\right)\left(\left[\hat{\mathbf{k}}\times \mathbf{B}_{\chi}\right]\cdot \bm{\nabla}_{\mathbf{k}_{\parallel}}\right)  \right\} (f_0-f_0^{\dag})
-\frac{iev_F^2}{\mu}
\left(\mathbf{A}_{\chi}\cdot\bm{\nabla}_{\mathbf{R}}\right) (f_0-f_0^{\dag}) \nonumber\\
&+&\frac{iev_F^2}{\mu}\left(\mathbf{A}_{\chi}\cdot\hat{\mathbf{k}}\right) \left(\hat{\mathbf{k}}\cdot\bm{\nabla}_{\mathbf{R}}\right) (f_0-f_0^{\dag})
+D_{\rm terms}^{(2)}
\Bigg\}.
\end{eqnarray}
Here, we used the explicit form of $\tilde{M}_{jl}$ in Eq.~(\ref{Eilenberger-sol-inter-M}) and introduced the following shorthand notations for the terms with the gap function derivatives:
\begin{eqnarray}
\label{Eilenberger-Sol-inter-Second-eq-Dterms1}
D_{\rm terms}^{(1)}&\equiv&-\frac{1}{4} \left[\left(\partial_{R_{j}} \partial_{R_{l}} \Delta_{\chi}\right) \left(\partial_{k_{\parallel, j}}\partial_{k_{\parallel, l}}f_0^{\dag}\right)
-\left(\partial_{R_{j}} \partial_{R_{l}} \Delta_{\chi}^{*}\right) \left(\partial_{k_{\parallel, j}}\partial_{k_{\parallel, l}}f_0\right)\right] \nonumber\\
&+&\frac{1}{2} \left[\left(\partial_{R_{j}} \partial_{k_{\parallel, l}}  \Delta_{\chi}\right) \left(\partial_{k_{\parallel, j}}\partial_{R_{l}}f_0^{\dag}\right)
-\left(\partial_{R_{j}} \partial_{k_{\parallel, l}}  \Delta_{\chi}^{*}\right) \left(\partial_{k_{\parallel, j}}\partial_{R_{l}}f_0\right)\right] \nonumber\\
&-&\frac{1}{4} \left[ \left(\partial_{k_{\parallel, j}}\partial_{k_{\parallel, l}} \Delta_{\chi}\right) \left(\partial_{R_{j}} \partial_{R_{l}}f_0^{\dag}\right)
-\left(\partial_{k_{\parallel, j}}\partial_{k_{\parallel, l}} \Delta_{\chi}^{*}\right)
\left(\partial_{R_{j}} \partial_{R_{l}}f_0\right)\right]
\end{eqnarray}
and
\begin{eqnarray}
\label{Eilenberger-Sol-inter-Second-eq-Dterms2}
D_{\rm terms}^{(2)}&\equiv&
-\frac{1}{4} \left[\partial_{R_{j}} \partial_{R_{l}} (\Delta_{\chi}+\Delta_{\chi}^{*})\right] \left(\partial_{k_{\parallel, j}}\partial_{k_{\parallel, l}}g_0\right)
+\frac{1}{2} \left[\partial_{R_{j}} \partial_{k_{\parallel, l}} (\Delta_{\chi}+\Delta_{\chi}^{*})\right]
\left(\partial_{k_{\parallel, j}}\partial_{R_{l}}g_0\right) \nonumber\\
&-&\frac{1}{4} \left[ \partial_{k_{\parallel, j}}\partial_{k_{\parallel, l}} (\Delta_{\chi}+\Delta_{\chi}^{*})\right]
\left(\partial_{R_{j}} \partial_{R_{l}}g_0\right).
\end{eqnarray}
In order to determine the remaining components of function $\tilde{g}_2$, we need to employ the
normalization condition (\ref{Model-Gor'kov-Eilenberger-g-norm-Eq}), which in the second order gives
\begin{eqnarray}
\label{Eilenberger-Sol-inter-Second-norm-1}
&&\bar{g}_2 = -g_2,\\
\label{Eilenberger-Sol-inter-Second-norm-2}
&&2g_0g_2 -f_2f_0^{\dag} -f_0f_2^{\dag} -f_2f_0^{\dag} -f_1f_1^{\dag}+g_1^2 =0.
\end{eqnarray}
Similarly to the first order case discussed in Sec.~\ref{sec:Eilenberger-sol-inter-First}, the explicit expressions for the components of $\tilde{g}_2$ can now be easily obtained, but
they are rather bulky. For studying the gap generation, however, it is sufficient to have only the
anomalous part $f_2$ of the integrated Green's function, i.e.,
\begin{eqnarray}
\label{Eilenberger-Sol-inter-Second-sol-f2}
f_2 &=& -i\frac{v_F f_0}{2\Delta_{\chi}}  \left(f_0 - \frac{2i\omega_{m^{\prime}}g_0}{\Delta_{\chi}+\Delta_{\chi}^{*}}\right) \left\{e\left(\left[\hat{\mathbf{k}}\times\mathbf{B}_{\chi}\right]\cdot \bm{\nabla}_{\mathbf{k}_{\parallel}}\right) -(\hat{\mathbf{k}}\cdot\bm{\nabla}_{\mathbf{R}})\right\}g_1
\nonumber\\
&-&i\frac{e^2v_F^2 f_0}{2\mu \Delta_{\chi}} \left(f_0 - \frac{2i\omega_{m^{\prime}}g_0}{\Delta_{\chi}+\Delta_{\chi}^{*}}\right) \left\{\left(\left[\mathbf{A}_{\chi}\times \mathbf{B}_{\chi}\right]\cdot \bm{\nabla}_{\mathbf{k}_{\parallel}}\right) -\left(\mathbf{A}_{\chi}\cdot \hat{\mathbf{k}}\right)\left(\left[\hat{\mathbf{k}}\times \mathbf{B}_{\chi}\right]\cdot \bm{\nabla}_{\mathbf{k}_{\parallel}}\right) \right\} g_0 \nonumber\\
&+&i\frac{ev_F^2f_0}{2\mu \Delta_{\chi}} \left(f_0 - \frac{2i\omega_{m^{\prime}}g_0}{\Delta_{\chi}+\Delta_{\chi}^{*}}\right) \left\{\left(\mathbf{A}_{\chi}\cdot\bm{\nabla}_{\mathbf{R}}\right) -\left(\mathbf{A}_{\chi}\cdot\hat{\mathbf{k}}\right) \left(\hat{\mathbf{k}}\cdot\bm{\nabla}_{\mathbf{R}}\right)\right\}g_0
-\frac{f_0}{4\Delta_{\chi}} \left(f_0 - \frac{2i\omega_{m^{\prime}}g_0}{\Delta_{\chi}+\Delta_{\chi}^{*}}\right) D^{(1)}_{\rm terms}
\nonumber\\
&+&\frac{f_0g_0}{2\left(\Delta_{\chi}+\Delta_{\chi}^{*}\right)} D^{(2)}_{\rm terms} +\frac{iv_F f_0g_0}{2\left(\Delta_{\chi}+\Delta_{\chi}^{*}\right)} \left\{e \left(\left[\hat{\mathbf{k}}\times\mathbf{B}_{\chi}\right] \cdot\bm{\nabla}_{\mathbf{k}_{\parallel}}\right) -(\hat{\mathbf{k}}\cdot\bm{\nabla}_{\mathbf{R}})\right\}(f_1-f_1^{\dag})
\nonumber\\
&+&\frac{ie^2v_F^2 f_0g_0}{2\mu\left(\Delta_{\chi}+\Delta_{\chi}^{*}\right)} \left\{\left(\left[\mathbf{A}_{\chi}\times \mathbf{B}_{\chi}\right]\cdot \bm{\nabla}_{\mathbf{k}_{\parallel}}\right) -\left(\mathbf{A}_{\chi}\cdot \hat{\mathbf{k}}\right)\left(\left[\hat{\mathbf{k}}\times \mathbf{B}_{\chi}\right]\cdot \bm{\nabla}_{\mathbf{k}_{\parallel}}\right) \right\} \left(f_0 \frac{\Delta_{\chi}-\Delta_{\chi}^{*}}{\Delta_{\chi}}\right) \nonumber\\
&-&\frac{iev_F^2 f_0g_0}{2\mu\left(\Delta_{\chi}+\Delta_{\chi}^{*}\right)} \left\{\left(\mathbf{A}_{\chi}\cdot\bm{\nabla}_{\mathbf{R}}\right) -\left(\mathbf{A}_{\chi}\cdot\hat{\mathbf{k}}\right) \left(\hat{\mathbf{k}}\cdot\bm{\nabla}_{\mathbf{R}}\right)\right\} \left(f_0 \frac{\Delta_{\chi}-\Delta_{\chi}^{*}}{\Delta_{\chi}}\right)
+\frac{f_0f_1f_1^{\dag}}{2} -\frac{f_0g_1^2}{2}.
\end{eqnarray}
The expression for $f_2^{\dag}$ can be obtained from $f_2 $ by using Eq.~(\ref{Eilenberger-Sol-inter-Second-eq-fdag2}).
Then, the expressions for the functions $g_2$ and $\bar{g}_2$ follow from Eqs.~(\ref{Eilenberger-Sol-inter-Second-eq-g2-gbar2}), (\ref{Eilenberger-Sol-inter-Second-norm-1}), and (\ref{Eilenberger-Sol-inter-Second-norm-2}).
As for the first order terms, they are given in Sec.~\ref{sec:Eilenberger-sol-inter-First}.
We note that the function $f_2$ in Eq.~(\ref{Eilenberger-Sol-inter-Second-sol-f2}) contains both odd and even terms with respect
to the Matsubara frequency.

\end{document}